\documentclass{eas}
\usepackage{graphicx}
\def\lesssim{\;\raise0.3ex\hbox{$<$\kern-0.75em\raise-1.1ex\hbox{$\sim$}}\;}
\def\gtrsim{\;\raise0.3ex\hbox{$>$\kern-0.75em\raise-1.1ex\hbox{$\sim$}}\;}
\newcommand{\beq}{\begin{equation}}
\newcommand{\eeq}{\end{equation}}
\newcommand{\bea}{\begin{eqnarray}}
\newcommand{\eea}{\end{eqnarray}}
\def\mdens{{\rm g~cm^{-3}}}
\def\bdens{{\rm fm^{-3}}}
\begin{document}
\title{EQUATION OF STATE OF DENSE MATTER AND MAXIMUM MASS OF NEUTRON STARS}
\author{ P. Haensel}
\address{N. Copernicus Astronomical Center, Bartycka 18,\\
00-718 Warszawa, Poland, {\tt haensel@camk.edu.pl}}
\runningtitle{Haensel: EOS and maximum mass of neutron stars}
\begin{abstract}
Theoretical models of the equation of state (EOS)
of neutron-star matter (starting with the crust
and ending at the densest region of the stellar core) are reviewed. Apart from
a broad set of baryonic EOSs, strange quark matter, and even more exotic (abnormal
and Q-matter) EOSs are considered. Results of calculations of $M_{\rm max}$
for non-rotating neutron stars and exotic compact stars are reviewed, with
particular emphasis on the dependence on the dense-matter EOS. Rapid rotation
increases $M_{\rm max}$, and this effect is studied for both neutron stars and exotic
stars. Theoretical results are then confronted with measurements of masses of neutron
stars in binaries, and the consequences of such a confrontation and their possible impact
on the theory of dense matter  are discussed.
\end{abstract}
\maketitle
\section{Introduction}
\label{sect:introd}
Since the pioneering paper of Oppenheimer and Volkoff (1939) the
problem of the actual value of the maximum mass of neutron stars,
$M_{\rm max}$, is puzzling both the theorists and observers.
Oppenheimer and Volkoff, who used free neutron-gas model, obtained
$M_{\rm max}=0.72~{\rm M}_\odot$. The precisely known mass
of the Hulse-Taylor pulsar is a direct proof of the dominating
role of the {\it nuclear forces} in the equation of state (EOS) of
neutron stars, which stiffen the EOS so that $M_{\rm
max}>1.44~{\rm M}_\odot$. Unfortunately, the value of $M_{\rm
max}$ is determined by the EOS of dense matter at $\rho>2\rho_0$,
where $\rho_0=2.7\times 10^{14}~\mdens$ is normal nuclear density,
and at such high densities  the EOS is largely unknown. This
results in a large uncertainty in theoretically derived value of
$M_{\rm max}$.

Within General Relativity, a compact  object of
$10~{\rm M}_\odot\gtrsim M>1~{\rm M}_\odot$
and $R<50$ km cannot be but a neutron star {\it or} a black hole.
All compact objects with $M>M_{\rm max}$ are necessarily black holes,
while only those with $M<M_{\rm max}$ can be neutron stars.
 From the observational point of view, the problem consists in determining
the {\it mass function} of neutron stars: its upper edge is just
$M_{\rm max}$. However, the present measurements of neutron stars masses
are most probably strongly biased due to a  specific evolutionary
formation scenario
(binary radio pulsars) or are not  very useful due to
large errors (X-ray binaries; some possible exceptions
are discussed in Sect.\ \ref{sect:observ}).
Hopefully, the knowledge of neutron-star
masses will improve in the future (see discussion in
Sect. \ref{sect:observ}). Then, confrontation of measured
masses with theoretical models will give us precious information on
the EOS of matter at $\rho\sim 10^{15}~\mdens$.

The present paper reviews the status of the theoretical determination
of $M_{\rm max}$. Sections \ref{sect:EOScrust}--\ref{sect:EOSabnormal}
are devoted to a brief description of  existing
models of the EOS of dense matter. Calculations of neutron-star structure
in General Relativity are presented in Sect.\ \ref{sect:NSmodels}. In
Sects.\ \ref{sect:Mrhoc.Mmax}-\ref{sect:stability}
 we discuss the mass--central-density diagram
for non-rotating (conventional) neutron stars made of baryons, with particular emphasis
on the value of $M_{\rm max}$ and its dependence on the EOS of baryonic matter,
as well as the stability of stellar configurations.
Short Sect.\ \ref{sect:Mmax.origin} is devoted to a discussion of the
origin of the maximum of neutron star mass. An upper bound on $M_{\rm max}$
is studied in Sect.\ \ref{subsect:struct-UBoundMmax}. The problem of
stability of stellar configurations is reviewed in Sect.\ \ref{sect:stability}.
Phase transitions in dense matter and their impact on neutron-star masses,
and in particular on $M_{\rm max}$, are described in
Sec.\ \ref{sect:struct-phase.tr}. Effect of rotation on neutron-star structure
and on the maximum mass is discussed in
Sects.\ \ref{sect:NSrot}-\ref{sect:Mmax.rot}. Section
\ref{sect:strange} is devoted to intriguing (hypothetical) family  of
strange quark stars, and Sect.\ \ref{sect:stars.abnormal} -- to even more
exotic Q-stars. Theoretical values of $M_{\rm max}$ are confronted with
existing measurements of neutron-star masses in Sect.\ \ref{sect:observ}.
Some final remarks are presented in Sect.\ \ref{sect:final}.
\section{Equation of state of the neutron-star crust}
\label{sect:EOScrust}
We assume that the crust is built of matter in full thermodynamic
equilibrium (the so called {\it cold catalyzed matter} corresponding to
minimum energy per nucleon at a given nucleon density);
the case of {\it accreted crust} will be briefly mentioned
at the end of the present section. The crust can
be divided into an {\it outer crust}, in which a lattice of nuclei is
immersed in an electron gas, and an {\it inner crust} consisting
of a lattice of nuclei immersed in an neutron gas and electron gas.
The outer crust ends at the neutron drip point $\sim 4\times 10^{11}~\mdens$,
while the inner crust extends down to the crust-core interface at
$\sim 10^{14}~\mdens$.

\begin{figure}
\begin{center}
\includegraphics[width=8cm]{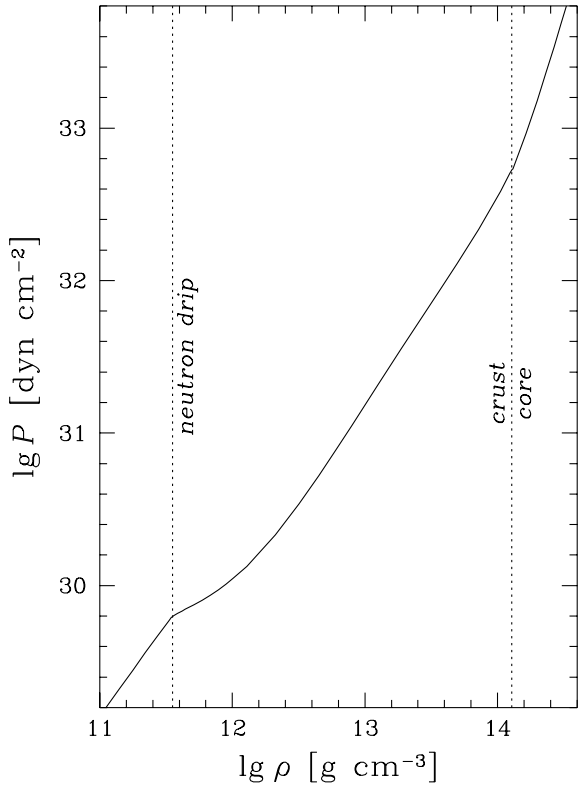}
\caption{
The SLy EOS (Douchin and Haensel 2001) of neutron star crust.
Dotted vertical lines
correspond to the neutron drip and crust-core transition.}
\label{fig:crust-SLyEOScrust}
\end{center}
\end{figure}
Up to the neutron drip point,  the EOS can be  calculated using the
experimental data on the neutron-rich nuclei and semiempirical nuclear
mass formulae  (Haensel and Pichon 1994).
Therefore, the  EOS of the outer crust is rather
well established. As soon as one leaves the region of experimentally
known nuclei, the EOS
of cold catalyzed matter becomes  uncertain. This uncertainty rises
at densities higher than the neutron drip density.
 The properties of nuclei are affected by the ambient neutron
gas which contributes more and more to the total pressure.
Therefore,
the problem of correct modelling of the EOS of pure neutron gas
at subnuclear densities becomes important. The real EOS of cold
catalyzed matter stems from the real nucleon Hamiltonian, which is expected
to describe nucleon  interactions at $\rho\lesssim 2\rho_0$.
  In practice, in order
to make the solution of the many-body problem feasible, the task is
reduced to  finding an {\it effective nucleon Hamiltonian}, which
would enable one to calculate reliably the EOS of cold catalyzed matter
for $10^{11}~\mdens \lesssim \rho \lesssim \rho_0$, including
the crust-core transition.

We will illustrate the properties of the inner crust EOS using the results
obtained by Douchin and Haensel (2001) for  the SLy model of
effective nuclear hamiltonian; it will be hereafter referred to as the
SLy EOS. Notice that the sound velocity is $v_{\rm s}=\sqrt{{\rm d}P/{\rm d}\rho}$.
The overall SLy EOS of the crust, calculated including
adjacent segments of the liquid core and
the outer crust, is shown in Fig. \ref {fig:crust-SLyEOScrust}.
In the outer crust segment, the SLy EOS cannot be visually distinguished
from the EOS  of Haensel and Pichon (1994).
 One notices a significant softening (decrease of $v_{\rm s}$)
just after the neutron drip point. At densities greater  than the
neutron drip one, the SLy EOS
stiffens gradually ($v_{\rm s}$ increases) with growing density,
due to the increasing contribution of dripped neutrons
to the pressure. There is  a discontinuous increase (jump)
of sound velocity at the crust-core interface.

In the SLy  EOS,  the crust-core transition takes place as
a very weak first-order phase transition, with the relative density jump
of the order of a percent (Douchin and Haensel 2001).
Let us remark  that for this model the spherical
nuclei persist to the crust bottom.
For the FPS EOS, the crust-core
transition takes place through a sequence of phase transitions with
the changes of nuclear shapes (spheres-rods-plates-tubes-bubbles,
 Lorenz et al. 1993).
 All in all, while the presence of the exotic nuclear
shapes is expected to have dramatic effect on  the transport
phenomena
 and elastic properties
of  matter, their effect on the EOS is small.

Some neutron stars, in particular those in close binary systems,
have crust composed of accreted matter. The EOS of {\it accreted crust}
is somewhat stiffer than for the ground-state crust (Haensel
and Zdunik 1990). However, the effect of this difference
 on the value of $M_{\rm max}$ is negligibly small.
\section{Equation of state of the neutron-star core}
\label{sect:EOScore}
\begin{table}[t]
\caption{Equations of state of the liquid core of neutron star}
\label{tab:struct-EOS}
\begin{center}
\begin{tabular}{|p{2cm}|p{6cm}|p{3cm}|}
\hline
\hline
EOS     & composition and model   & reference \\
\hline
\hline
{ BPAL12 }   & $npe\mu$, effective nucleon energy functional&
 Bombaci et al. 1995\\
\hline
{BGN1H1}  &  $np\Sigma\Lambda\Xi e\mu$, effective baryon energy
functional & Balberg et al. 1999\\
\hline
{BBB1}   & $npe\mu$, Brueckner theory, Argonne   NN
 plus  Urbana NNN potentials    &  Baldo  et al. 1997\\
\hline
{ FPS}   &  $npe\mu$, effective nucleon energy functional &
  Pandharipande and Ravenhall 1989 \\
\hline
{BGN2H1}  &  $np\Sigma\Lambda\Xi e\mu$, effective baryon energy
functional & Balberg et al. 1999\\
\hline
{BBB2}   & $npe\mu$, Brueckner theory, Paris NN
 plus  Urbana NNN  potentials &  Baldo  et al. 1997\\
\hline
{SLy}   & $npe\mu$, effective nucleon energy functional
 &  Douchin and  Haensel 2001\\
\hline
{BGN1}  &  $npe\mu$, effective baryon energy
functional & Balberg et al. 1999\\
\hline
{APR}   &  $npe\mu$, variational theory,
 Nijmegen NN plus  Urbana
 NNN potentials &
Akmal et al. 1998\\
\hline
{BGN2}  &  $np e\mu$, effective nucleon energy
functional   & Balberg et al. 1999\\
\hline
\hline
\end{tabular}
\end{center}
\end{table}
\begin{figure}
\begin{center}
\includegraphics[width=11cm]{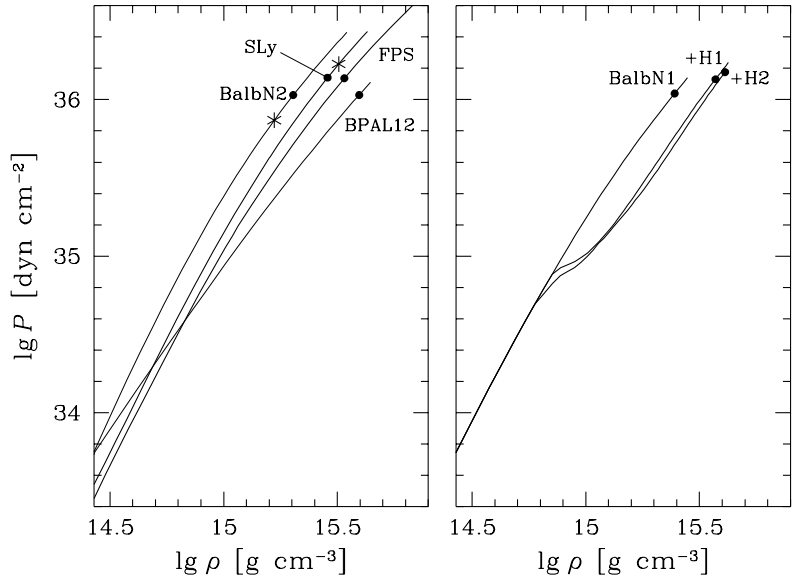}
\caption{
Selected models of the EOS of neutron star cores
denoted  as in Table \ref{tab:struct-EOS}, except for the BGN1
and BGN2 which here have labels BalbN1 and BalbN2.  Large dots
correspond to maximum density in stable neutron stars. Asterisks
correspond to the density above which EOS is superluminal
($v_{\rm s}>c$). Left panel: EOSs of the $npe\mu$ matter.
Right panel: effect of hyperons
is shown by comparing the  EOS without hyperons (i.e.,
for the $npe\mu$ matter)
and EOSs in which hyperons $\Lambda,~\Sigma,~\Xi$ are
included (+H1 and +H2 correspond
to the  BGN1H1 and BGN1H2 models of Table\ \ref{tab:struct-EOS}).
}
\label{fig:EOScore}
\end{center}
\end{figure}

\begin{figure}
\begin{center}
\includegraphics[width=9cm]{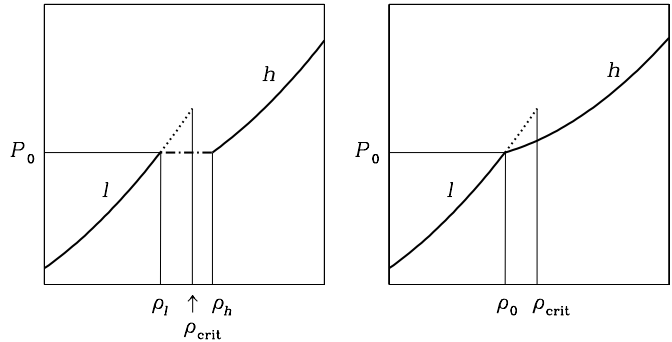}
\caption{
Effect of phase transitions on the EOS of neutron-star core. Solid line:
stable lower-density (l) or higher-density (h) phase.
Dotted line: metastable l-phase. The density at which 
metastable l-phase becomes unstable
with respect to the transition into a dense phase is denoted by
$\rho_{\rm crit}$. Left panel: first-order phase transition at $P=P_0$
from a pure l-phase to a pure h-phase, with the density jump at $P=P_0$
from $\rho_{\rm l}$ to $\rho_{\rm h}$. Right panel: transition from l-phase
to a mixed h-phase at $P=P_{\rm 0}$ and $\rho=\rho_{\rm 0}$. Notice that
here $\rho_0$ means the threshold density
for the equilibrium-phase transition.
The right panel could describe also the effect of
a second-order phase transition,
provided  the dotted segment is removed.
}
\label{fig:EOSpt}
\end{center}
\end{figure}
At $\rho\lesssim \rho_0$, core  matter is a liquid composed mostly
of neutrons with a  few percent admixture of the equal number of
protons and electrons. If the Fermi energy of electrons exceeds
the muon rest energy (105.7 MeV), muons replace a fraction of
electrons to minimize the energy of the system. Such a system in
beta equilibrium is usually called the $npe\mu$-matter. This is
the simplest model of matter in neutron star cores: except for the
presence of muons, which are insignificant for the EOS, the matter
constituents - neutrons, protons, and electrons - are the same as
in  familiar terrestrial matter. Still, even for this simplest
composition, the uncertainties in the EOS are quite large,
especially at densities significantly higher than $\rho_0$. This
results from the approximations and deficiencies of the
many-body theory of dense nucleon matter, and from the lack of
knowledge of strong interactions in superdense matter. We
illustrate these uncertainties in the left panel of Fig.\
\ref{fig:EOScore}, where four examples of the EOS of the
$npe\mu$-matter are shown. The brief characteristics of these
models and the references to the original papers are given in
Table \ref{tab:struct-EOS}. The BPAL12 and BGN2 EOSs should be
considered as the soft and stiff extremes of the theoretical
models. Only segments below large dots, which correspond to the
maximum density in stable neutron stars calculated for this EOS
(see Sect.\ \ref{sect:Mrhoc.Mmax} and in particular Table
\ref{tab:struct-EOSmax}),  are relevant for neutron stars. Notice
that the stiffest BGN2 EOS is superluminal (sound velocity $v_{\rm
s}>c$) at the highest densities relevant for neutron stars, which
reflects the inadequacy of the non-relativistic approach at such a
high density.

Let us denote the chemical potentials of the constituents of the $npe\mu$-matter
by $\mu_n$, $\mu_p$, $\mu_e$, and $\mu_\mu$. As soon as the sum $\mu_n+\mu_e$
exceeds the in-medium energy of $\Sigma^-$ hyperon, this
strange baryon will appear in as a stable constituent of
dense matter.
 $\Sigma^-$ is expected to be
the first hyperon to appear in dense matter at some $\rho_1$.
The second hyperon to appear, above a higher threshold density
$\rho_2$, is $\Lambda^0$.  Hyperons appear due to
strangeness-changing weak interactions.  At a given baryon density, the inclusion
of hyperons lowers significantly the matter pressure compared
to the case of the $npe\mu$ mater, because the highest-energy neutrons
are then replaced by the low-energy hyperons. This effect of
the softening of the EOS of dense matter due to  the presence
of hyperons is clearly seen in the right panel of Fig.\ \ref{fig:EOScore}.

Pion or kaon condensation in dense matter, as well as deconfinement of quarks,
predicted by some theories,
imply a softening of the EOS above the phase-transition threshold.  The effect
of such a phase transition on the EOS  is visualized in
Fig.\ \ref{fig:EOSpt}.
\section{EOS of strange  matter}
\label{sect:EOSstrange}
Strange quark matter (SQM)
is a hypothetical self-bound
state of deconfined quarks
which form a plasma of up (u), down (d), and strange (s) quarks with
$n_{\rm u}\simeq n_{\rm d}\simeq n_{\rm s}$  of total baryon number
density $n_{\rm b}=(n_{\rm u}+n_{\rm d}+n_{\rm s})/3$
and a small admixture of electrons
$n_e\sim (10^{-5}-10^{-4}) n_{\rm b}$.
The existence of a self-bound state at zero pressure results from the
pressure exerted by the ordinary vacuum on a  volume of the QCD vacuum
in which the quarks can  move. In the  Bag Model
 this pressure is equal to the bag constant $B\sim
60~{\rm MeV/fm^3}$. In the simplest Bag Model with massless free
quarks the density of SQM at zero pressure is $\rho_{\rm
s}=4B/c^2$. The hypothesis that SQM is the true ground state of
matter at zero pressure means that  energy per unit baryon number
$E_{\rm SQM}<E(^{56}{\rm Fe}) = 930.4~$MeV. Recent reviews of
physics of strange quark matter are  given by  Glendenning (1999),
Madsen (1999), and  Weber (1999).

As the pressure of SQM vanishes  at
$\rho_{\rm s}$ (larger than normal nuclear density  $\rho_0$),
the EOS has  a  linear form $P\simeq (\rho-\rho_{\rm s})\times
{\rm constant}$
 for densities only slightly higher than $\rho_{\rm s}$.
It is very fortunate that this simple linear form of the EOS turns out to be
an excellent approximation at higher densities, up to maximum density in stable
strange stars. This property is independent of the SQM model, and holds
for the MIT Bag Model (Zdunik 2000) and other  models of SQM (Gondek-Rosi{\'n}ska et al.
2000).

Let us focus our attention on the Bag Model. A linear approximation reads
\begin{equation}
P/c^2=a(\rho-\rho_{\rm s})~,
\label{strange-Prho.lin}
\end{equation}
where {\it constants} $a$ and $\rho_{\rm s}$ are determined by fitting the
exact EOS (Zdunik 2000). Notice that the linear form, Eq.\ (\ref{strange-Prho.lin}),
is exact in the case of massless quarks (free or interacting).
Using first law of thermodynamics one can easily see that Eq.\
(\ref{strange-Prho.lin}) implies
\begin{equation}
n_{\rm b}(P)=n_{\rm s}
\left[1 +
\left(4 -{1\over 3a}\right){P\over \rho_{\rm s}c^2}
\right]^{1/(a+1/3)}~,
\label{strange-nP.lin}
\end{equation}
where $n_{\rm s}$ is the baryon density of SQM at zero pressure.
Therefore three fitting parameters --  $\rho_{\rm s}$, $n_{\rm s}$, and $a$
-- completely determine  an EOS of SQM.
\section{EOS of  abnormal and Q-matter}
\label{sect:EOSabnormal}
Several  types of hypothetical self-bound state of dense baryon
matter, which could
constitute true ground state of the matter at $P=0$, were suggested in the past. In
1970s, it was suggested that the pion condensation could lead to the appearance of a
self-bound, superdense  state of pion-condensed nucleon matter (Migdal 1971, 1974,
Hartle et al. 1975). This phase of matter was expected to  form  ``abnormal
nuclei'' of large $A$ and with
the density significantly higher than $\rho_0$ (Migdal 1971,
1974) but   no sign of experimental evidence of the
existence of the ``abnormal nuclei'' has
been   found up to the time of this writing.

The idea of ``abnormal state'' of nuclear matter, as advanced by
Lee and Wick (1974; for a review see Lee 1975), is based on a
schematic field-theoretical  $\sigma$-model of strongly interacting
nucleon matter. In the abnormal state, which appears at
sufficiently high nucleon densities, the nucleons are thought to
become nearly massless.
This is because the $\sigma$-field term couples to
the nucleons as an (negative) addition to the nucleon rest (bare)
mass, implying a nearly vanishing nucleon effective mass.
This density-dependent effect
can lead to the appearance of a second minimum in the
$n_{\rm b}$ dependence of the
energy per baryon at some $n_{\rm a}$,   significantly
higher than $n_0$, with the binding energy significantly larger than
in  the ``normal state'' at $n_{\rm b}=n_0$.
 However, the original $\sigma$-model
is schematic and does not pretend to describe quantitatively
the normal nuclear matter at  $n_{\rm b}\approx
n_0$. Were the $\sigma$-model
more complicated to give correct quantitative description of
the nuclear
matter at $n_{\rm b}\approx n_0$, then the abnormal state at supranuclear
density would disappear (Pandharipande and Smith 1975, Moszkowski and
K{\"a}llman 1977).

Supersymmetric extensions of the Standard Model of elementary
particles and their interactions predict the existence of
macroscopic self-bound superdense lumps of a scalar-field
condensate with a well defined electric and baryonic charge --
baryonic Q-balls. The Q-balls were proposed as a hypothetical
component of cosmological dark matter (Kusenko et al. 1998,
Kusenko 2000). It was  also proposed
that Q-balls of a macroscopic size could have energy per unit  baryon
number lower than $^{56}{\rm Fe}$.
EOSs of Q-matter  were constructed by Bahcall et al.
(1990). A common feature of these models,
  shared   with the Lee-Wick model of abnormal matter is
that nucleons become massless inside the condensed Q-phase. Two
basic parameters of the model are: the
energy density  $U_0$ of the scalar
field inside the Q-matter,  and
the coupling strength $\alpha_v$ of the vector field to
the nucleon.
It is convenient to introduce a  dimensionless
parameter $\zeta=\alpha_v U^{1/2}_0\pi/\sqrt{3}$,
and to use it instead of the parameter $\alpha_v$.

Consider the simplest case
of $\zeta=0$. The standard value used
 by  Bahcall et al. (1990) is $U_0=13.0~{\rm
MeV~fm^{-3}}$, which  corresponds to $\rho(P=0)\equiv
\rho_{\rm s}=1.0\times
10^{14}~\mdens$: for this model the density of Q-balls is
subnuclear.
With increasing $\zeta$, the EOS of the Q-matter becomes stiffer
and the value of $\rho_{\rm s}$  lower. In
the limiting case of $\zeta=16$  considered by  Bahcall et al. (1990)
(at the same value of $U_0=13.0~{\rm MeV}~\bdens$)  they get
$\rho_{\rm s}
=5.5\times 10^{13}~\mdens$.
 The predicted density of self-bound Q-matter
at zero pressure is two to five times lower (!) than the normal nuclear density, which
results from a strong reduction of effective nucleon masses in this hypothetical
state of nucleon matter.

In all cases considered in the present section, the EOS of an
abnormal phase of dense
baryonic matter can be well approximated by
\begin{equation}
{\rm abnormal~matter,~Q-matter:}
~~~~ P\simeq a(\rho - \rho_{\rm s})c^2~.
\label{EOS.abnormal}
\end{equation}
This EOS is usually stiffer than that of strange quark matter,
and one  often has $a\simeq 1$.

Finally, let us stress that there has been no observational evidence for
the existence of hypothetical self-bound states of dense matter reviewed in the
present section  by the time of this writing.
\section{Neutron star models}
\label{sect:NSmodels}
Neutron stars are relativistic objects. Their structure and
evolution should be  studied using the General Theory of
Relativity. The importance of relativistic effects for a star of
mass $M$ and radius $R$ is characterized by the {\it compactness
parameter} $r_{\rm g}/R$, where $r_{\rm g}=2GM/c^2=2.95 \,(M/{\rm
M}_\odot)$ km is the gravitational radius. For a non-rotating
black hole, $r_{\rm g}/R=1$. Typically, one has $r_{\rm g}/R \sim$
0.2 -- 0.4 for a neutron star, while $r_{\rm g}/R \ll 1$ for all
other stars, e.g., $r_{\rm g}/R\sim 10^{-4}$ for white dwarfs, and
$r_{\rm g}/R\sim 10^{-6}$ for main-sequence stars of $M \simeq
1~{\rm M_\odot}$.

 We will approximate the stress tensor
 of neutron-star matter by that of a perfect fluid, i.e.
non-viscous medium of total energy density ${\cal E}$ (mass density
$\rho={\cal E}/c^2$), in which
all stresses are zero except for an isotropic pressure $P$. The
perfect-fluid approximation is justified by the fact that the
shear stresses, e.g., those produced by elastic strain in the
solid crust or by strong magnetic field, are generally negligible
compared to the pressure.

The problem of finding hydrostatic equilibrium
simplifies considerably in the case of strongly
degenerate neutron star interior where  thermal contributions to
$P$ and $\rho$ can be neglected. Then, the EOS
of matter involves only $P$ and $\rho$, but not the temperature,  a
situation which is nearly always valid in neutron-star interior
(important exceptions are: neutron-star atmospheres, newly-born
neutron stars, envelopes of exploding X-ray bursters).

For a static spherically symmetric neutron-star configuration
the equation of hydrostatic equilibrium  is (see,
e.g., Shapiro and Teukolsky 1983)
\begin{eqnarray}
   {{\rm d}P \over {\rm d}r} & = &
     - {G \rho m \over r^2 } \left( 1 + {P \over \rho c^2} \right)
     \left( 1 + {4 \pi P r^3 \over m c^2 } \right)
     \left( 1 - {2 G m \over c^2 r} \right)^{-1},
\label{str-P} \\
   {{\rm d}m \over {\rm d}r} & = & 4 \pi r^2 \rho,
\label{str-m}
\end{eqnarray}
where $m(r)$ is the  mass  contained within a
sphere of (coordinate) radius $r$.
 Equation (\ref{str-m}) describes mass balance; its
apparently Newtonian form is illusive since the space-time is curved.
 Equations (\ref{str-P})--(\ref{str-m})
 should be supplemented by an EOS, $P=P(\rho)$. The above
equations constitute then a closed system of equations to be solved for
obtaining $P(r)$, $\rho(r)$ and  $m(r)$.

In the stellar interior, $P>0$ and ${\rm d}P/{\rm d}r<0$. The
stellar radius is determined from the condition $P(R)=0$. Outside
the star, i.e., for $r
> R$, we have  $P=0$ and $\rho=0$, and from Eq.\ (\ref{str-m}) we obtain
$m(r>R) = M = $ const. The latter quantity is called the {\it
total gravitational mass of the star}; the total energy content of
a star is $E=Mc^2$.

 In order to construct a neutron-star model one has
to solve the set of Eqs.\ (\ref{str-P})--(\ref{str-m}) supplemented
by an EOS of neutron-star matter.
Let us notice that  the mass density and baryon density are
 not constrained to be continuous. In
principle,  density discontinuities can appear at pressures
corresponding to first-order phase transitions in dense matter
(see Sect.\ \ref{sect:EOScore}).
 The density profile
is  determined using  the EOS, through $\rho=\rho(P)$.

Equations (\ref{str-P})--(\ref{str-m}) are integrated
from the stellar center with the following boundary conditions:
$P(0)=P_{\rm c}$, $m(0)=0$. The stellar surface, $r=R$,
is determined from $P(R)=0$.

 EOS constitutes a crucial input for the neutron-star structure calculations.
Models of the EOS were described in Sects.\ \ref{sect:EOScrust}--\ref{sect:EOScore}.
Here we briefly describe the EOS used in  neutron-star structure calculations
discussed  in the present paper.
The outer neutron-star crust is
described by the EOS of Haensel and Pichon (1994).
 The inner crust is
described by the FPS model, developed by Lorenz
(1991), except for the SLy  EOS, which
describes in a unified manner both
the core and the crust.
Actually, a
particular choice of the crustal EOS has little importance for the
global parameters of massive neutron-star models, which are mostly
determined  by the EOS of the liquid core at  supranuclear
densities.

The EOS of the core is dominated by the contributions from {\it
strong (nuclear) interactions} between nucleons (or, more
generally, between baryons; see Sect.\ \ref{sect:EOScore}). The
reliability of the theory of a dense, strongly interacting
many-body system decreases rapidly with increasing $\rho$.
Moreover, we have very little experimental information about
baryonic interactions at supranuclear densities, especially about
the interactions  involving hyperons which may appear
 at higher densities. As a consequence, the EOS
of neutron-star matter at supranuclear densities is strongly
dependent on the specific microscopic theory of dense matter
employed.

Our discussion of the impact of the EOS$(\rho>\rho_0)$ on the
neutron-star structure  will be based on the  limited, but
hopefully representative, set of models of dense {\it baryonic}
matter.\footnote[1]{Stellar models based on the EOSs including
hypothetical exotic phases, such as pion condensate, kaon
condensate, and deconfined quark matter, will be considered
separately in Sect. \ref{sect:struct-phase.tr}. A specific case of
hypothetical {\it strange (quark) stars} or even stranger stars
built of {\it abnormal matter} or {\it Q-matter} will be briefly
reviewed in Sect.\ \ref{sect:strange} and
\ref{sect:stars.abnormal}. }
The selected  models of dense cold baryonic matter are listed in
Table \ref{tab:struct-EOS}. They were briefly discussed in
Sect.\ \ref{sect:EOScore}.
As we stressed in Sect.\ \ref{sect:EOScore}, the most important
qualitative feature of an  EOS at $\rho>\rho_0$ -- as far as
neutron-star structure is concerned -- is its stiffness. A simple
definition of the stiffness  can be phrased as follows: the
higher the value of $P$ at a given $\rho$, the stiffer the EOS.
However, this definition is ambiguous and sometimes misleading. A
somewhat better definition relies on the dimensionless adiabatic
index $\gamma=(n_{\rm b}/P)({\rm d}P/{\rm d}n_{\rm b})$. However,
a core-EOS
relevant for the complete family of neutron-star models has to be
given for $\rho_0\lesssim \rho \lesssim (10-15)\rho_0$. It can be
rather  stiff at $\rho \lesssim (2-3)\rho_0$, and then become softer  at
higher $\rho$, due to the hyperonization of matter with increasing
density (see Sect.\ \ref{sect:EOScore}).  Therefore, it is  useful to
base the {\it effective stiffness} of the EOS on the value of the
maximum allowable mass of  neutron stars, $M_{\rm max}$, which is
a functional of the EOS. The topmost EOS in Table
\ref{tab:struct-EOS} is the softest one, and the effective
stiffness of the EOSs increases from the top of the table to its
bottom (see Table \ref{tab:struct-EOSmax}). Let us remind, that
the stiffest/softest EOSs are extreme models, characterized by the
compression modulus of symmetric nuclear matter, which is
significantly higher/lower than the standard ``empirical value''.
These
EOSs are  included to represent the  theoretical
extremes of the EOS
of dense matter.

Two of the EOSs listed in Table \ref{tab:struct-EOS}, namely SLy
and FPS,  are the {\it unified} ones, and  describe,  within  a
single physical model,  both the crust and  the core of a
star (see Sect.\ \ref{sect:EOScrust}). In all other cases, an
overall EOS of the stellar  interior was constructed by matching
smoothly the EOS of the crust with that of the core.
\begin{figure}
\begin{center}
\includegraphics[width=8cm]{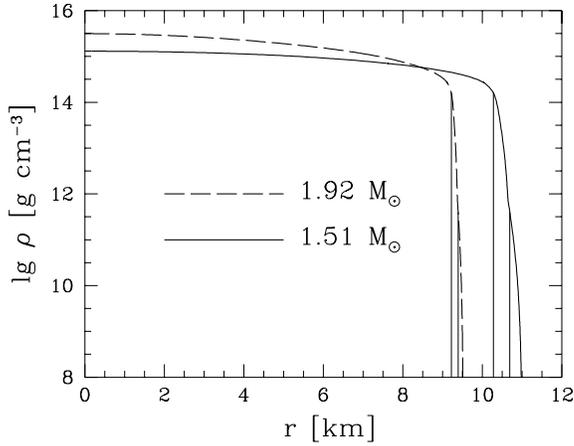}
\caption{
Mass density versus radial coordinate for neutron-star models
of gravitational mass
 $1.51~{\rm M}_\odot$ (solid line) and $1.92~{\rm
M}_\odot$ (long dash-dotted line). Calculations are performed for the BBB2 EOS of the
core, for which $1.92~{\rm M}_\odot$ is the maximum allowable mass. The positions  of
the crust-core interface and the neutron-drip point  are indicated by thin vertical
lines.}
 \label{fig:str-prof3M}
 \end{center}
\end{figure}

\begin{figure}
\begin{center}
\includegraphics[width=8cm]{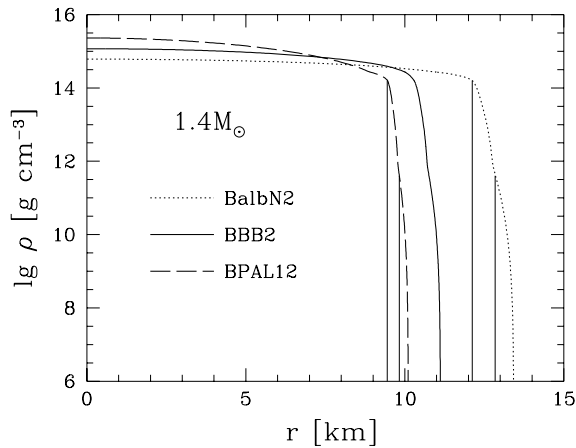}
\caption{
Mass density versus radial coordinate for neutron-star models
of gravitational mass $1.4~{\rm M}_\odot$, for the core  described by the BPAL12 EOS
(long dash-dotted line), BBB2 EOS (solid line), and BalbN2 
( denoted as BGN2 in Table \ref{tab:struct-EOS}) 
EOS (dotted line). For the BPAL12 and
BalbN2 
models, positions of the crust-core interface and the neutron-drip point
 are indicated by thin vertical
lines.   }
 \label{fig:str-prof1.4}
 \end{center}
\end{figure}
\subsection{Basic internal structure of neutron stars}
\label{sect:int.struct}
The distribution of matter within a neutron star depends on its
mass and results from an interplay between the pressure and
gravity. In Fig.\ \ref{fig:str-prof3M} we show the mass density
within a neutron star as a function of radial coordinate,
calculated for the BBB2  EOS of the core, for two  masses. The
higher  mass is equal to the maximum  mass allowable for this EOS.
The calculations show that the density within the core is rather
uniform, for $M=(1.2-1.5)~{\rm M}_\odot$; the most pronounced drop
occurs  close to the crust bottom edge. There is a steepening of
the density drop near  the neutron drip point, which results from
the softening of the EOS due to the neutron drip. The increase of
the stellar mass increases the gravitational pull within the
crust, which leads to the steepening of density profiles and
thinning of the crust, most pronounced  at $M=M_{\rm max}$.  For a
medium-stiff BBB2  EOS of the  core,  the  crust contains $1.4\%$
and $0.8\%$  of the total gravitational mass of $1.24~{\rm
M}_\odot$ and $1.51~{\rm M}_\odot$ stars,  and the crust thickness
is  $1.01$ km and $0.72$ km, respectively.
 At the maximum mass, the crust contains
 only $0.2\%$ of the stellar mass, and the crust
thickness reduces to about $0.29~$km. The crust mass and
thickness  are smaller for softer EOSs  of the core, and
larger for  stiffer ones.

 While the crustal EOS  is rather well established (see
Sect.\ \ref{sect:EOScrust}), the crust structure results from the
interplay of its EOS and the gravitational pull  exerted by
the core: the latter depends on the core compactness. Therefore,
the uncertainty of  the EOS in the  core implies  uncertainty in
the crust structure. This is visualized in Fig.\
\ref{fig:str-prof1.4}, where we show the mass  density profiles
inside a 1.4~${\rm M}_\odot$ star calculated for the BPAL12, BBB2,
and BGN2  EOSs of the core. The crust thickness ranges from 0.7 km
for the softest core EOS up to 1.3 km for  the stiffest one. The
dependence
  of  the fraction of stellar
mass contained in the crust on the core EOS
is even more dramatic: this fraction ranges
from 0.7\% for the softest BPAL12 EOS to
 2.2\% for the stiffest BGN2  one.
The BBB2
model is   typical for medium-stiff EOSs: the crust
thickness is about  0.8 km and the mass  fraction
is 1\%, respectively.
\subsection{Mass--central-density diagram and $M_{\rm max}$}
\label{sect:Mrhoc.Mmax}
\begin{figure}
\begin{center}
\includegraphics[width=10cm]{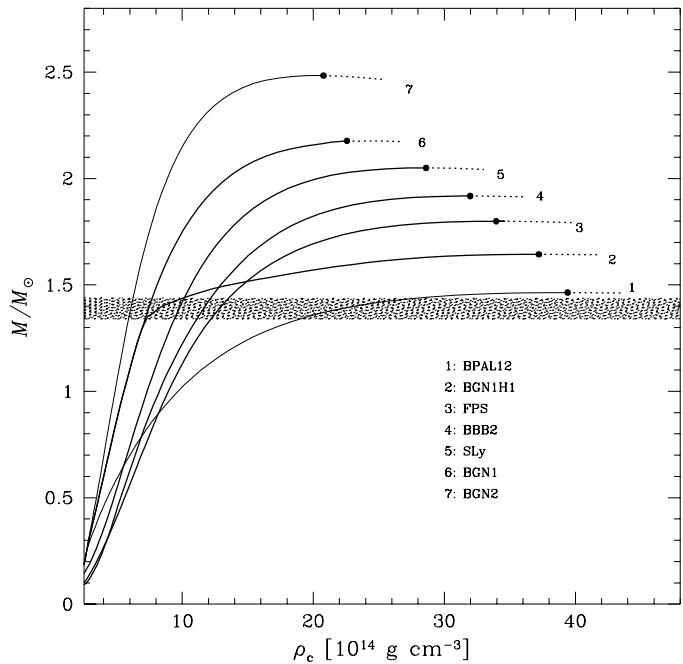}
\caption{
Gravitational mass $M$ versus central
density $\rho_{\rm c}$ of static neutron-star models for
different EOSs. A maximum  on a  $M-\rho_{\rm c}$ curve is
indicated by a filled circle. Configurations to the right of the maximum are
unstable with respect to small radial perturbations, and are
denoted by dotted lines. The shaded band corresponds to the range
of precisely  measured masses of binary radio pulsars (Sect.\ \ref{sect:observ}).
}
 \label{fig:str-Mrhoc}
 \end{center}
\end{figure}

The models of cold, static neutron stars  form a one-parameter
family; they can be labelled by their  central density $\rho_{\rm
c}$, or central pressure $P_{\rm c}$. In Fig.\ \ref{fig:str-Mrhoc}
we show dependence of the gravitational mass, $M$, on $\rho_{\rm
c}$ for some  EOSs from Table \ref{tab:struct-EOS} at $\rho_{\rm
c}>2.5\times 10^{14}~{\rm g/cm^3}$.

\begin{table}[t]
\caption{Maximum mass configurations for EOSs listed in Table
\ref{tab:struct-EOS}. Displayed parameters are:  radius $R$,
compactness parameter $r_{\rm g}/R$, central baryon density
$n_{\rm c}$,  and central mass density $\rho_{\rm c}$. }
\label{tab:struct-EOSmax}
\begin{center}
\begin{tabular}{|p{2cm}|p{1cm}|p{1cm}|p{1cm}|p{1cm}|p{2.5cm}|}
\hline \hline EOS     &  $M_{\rm max}$ &  $R$ & $r_{\rm g}/R$ &
$n_{\rm c}$ & $\rho_{\rm c}$
\\
    &  $[M_{\odot}]$ &  [km]  &  & $[{\rm fm^{-3}}]$ &
    $[{\rm 10^{15}~{g~cm^{-3}}}]$ \\
\hline \hline
 BPAL12   &  1.46   & 9.00  & 0.480  &  1.76 & 3.94\\
 BGN1H1 &  1.64   & 9.38  & 0.519  &  1.60 & 3.72\\
 BBB1   &  1.79   & 9.66  & 0.547  &  1.37 & 3.09\\
 FPS   &  1.80   & 9.27  & 0.572  & 1.46 & 3.40\\
 BGN2H1 &  1.82   & 9.53  & 0.564  &  1.45 & 3.48\\
 BBB2   &  1.92   & 9.49  & 0.596  &  1.35 & 3.20\\
 SLy  &   2.05   & 9.99  & 0.605  &  1.21 & 2.86\\
 BGN1 &  2.18   & 10.9  & 0.592  &  1.05 & 2.26\\
 APR   &  2.21 &    10.0   & 0.651  & 1.15 & 2.73\\
 BGN2 &  2.48   & 11.7  & 0.626  &  0.86 & 2.02\\
\hline \hline
\end{tabular}
\end{center}
\end{table}

On the lower-density side, the $M(\rho_{\rm c})$ curves exhibit a minimum
$M_{\rm min}\simeq 0.1~{\rm M}_\odot$, which depends rather
weakly on the assumed EOS (Haensel et al. 2001),
and which is not shown in Fig.\ \ref{fig:str-Mrhoc}.

On the higher-density side, the $M(\rho_{\rm c})$ curves exhibit a
maximum, which strongly depends  on the assumed EOS, and is
reached at densities a few times $10^{15}~{\rm g~cm^{-3}}$.
Configurations with $M>M_{\rm max}$ cannot exist in hydrostatic
equilibrium and must necessarily collapse into black holes. For
the selected EOSs $M_{\rm max}$  ranges from $1.46~{\rm M}_\odot$
to $2.48~{\rm M}_\odot$~ (Table \ref{tab:struct-EOSmax}).

 Replacing
neutron-star matter by a free (noninteracting) Fermi gas of
neutrons lowers the value of $M_{\rm max}$ to $0.72~{\rm M}_\odot$
(Oppenheimer and Volkoff 1939). If one further allows for the beta
equilibrium between otherwise noninteracting  nucleons, electrons,
and muons, the EOS becomes even softer, with $M_{\rm
max}=0.70~{\rm M}_\odot$. Clearly, the values of $M_{\rm max}$
obtained neglecting nuclear interactions are in
contradiction with precisely measured masses of the binary radio
pulsars (Sect.\ \ref{sect:observ}).
 Turning the argument around, we may say that the measured
 mass  $1.44~{\rm M}_\odot$ of the Hulse-Taylor pulsar
 (Sect.\ \ref{sect:observ}) implies that nucleon-nucleon interaction is
sufficiently repulsive at supranuclear densities to lift $M_{\rm
max}$ by more than hundred percent  from the non-interacting
nucleon  gas value.

In view of uncertainties in the hyperon-nucleon (H-N) and
hyperon-hyperon (H-H) interaction in dense matter, let us first
consider a subset of EOSs with ${\rm N}e\mu$ composition of
matter (N denotes $np$). Let us remove from this subset the extreme models BPAL12
and BGN2. Then the subset will be restricted to the models of the
${\rm N}e\mu$ matter which reproduce  empirical saturation properties of
nuclear matter (see Sect.\ \ref{sect:EOScore}). Under such
conditions, we narrow the range of theoretical maximum allowable
mass of  neutron stars,
%
\begin{equation}
M_{\rm max}({\rm Ne}\mu)\simeq (1.8-2.2)~{\rm M_\odot}~.
\label{struct-MmaxNemu}
\end{equation}
%
The appearance  of hyperons softens an EOS, and therefore
decreases the value of $M_{\rm max}$. For the selected EOSs, the
appearance of hyperons at $\rho\simeq 2\rho_0$ lowers the value of
maximum allowable mass to an even narrower range,
%
\begin{equation}
M_{\rm max}({\rm NH}e\mu)\simeq (1.5-1.8)~{\rm M_\odot}~.
\label{struct-MmaxNHemu}
\end{equation}
%
Let  us stress that this conclusion is obtained  under the
assumption that the hyperons  appear at about $2\rho_0$. However,
the threshold for the hyperon appearance depends sensitively on
the N-H interaction in dense matter, which is poorly known for a
neutron rich dense matter at $\rho \gtrsim \rho_0$. Therefore, the
values of $M_{\rm max}$ in Eq.\ (\ref{struct-MmaxNHemu}) have  to be
taken with a grain of salt. Actually, it is  reasonable to say
that Eq.\ (\ref{struct-MmaxNHemu}) corresponds to a typical effect
of hyperons on $M_{\rm max}$ (and on neutron-star structure),
assuming that without hyperons $M_{\rm max}$ is  given by Eq.\
(\ref{struct-MmaxNemu}).
Alas, the lack of knowledge of the H-H interaction implies
 large uncertainty in the effect of hyperons on $M_{\rm max}$.
 For example, an EOS calculated
by Vida$\tilde{\rm n}$a  et al. (2001) shows particularly  strong softening
 due to the
hyperonic interactions, with $M_{\rm max}({\rm N}e\mu)=1.89
~{\rm M}_\odot$ and $M_{\rm max}({\rm NH}e\mu)=1.34~{\rm M}_\odot$.
The last value contradicts the measured mass of the Hulse-Taylor
pulsar. When Vida$\tilde{\rm n}$a et al. (2001) remove (artificially)  the H-H
interaction they get $M_{\rm max}({\rm NH}e\mu)=1.47~{\rm M}_\odot$,
which is marginally acceptable.
%
\subsection{Stability of neutron stars}
\label{sect:stability}
The mass-central density diagram shows that cold, static
equilibrium configurations of neutron stars can exist only for
$M_{\rm min}<M<M_{\rm max}$. The value of $M_{\rm min}$ seems to
be rather well established, $M_{\rm min}\simeq 0.1~{\rm M}_\odot$ 
(see Haensel et al. 2002 and references therein).
Equilibrium configurations  of neutron stars with $M<M_{\rm min}$
do not exist.

Non-rotating
 configurations with $M>M_{\rm max}$ collapse into black
holes. Numerical simulations of such a collapse were
performed by several authors, e.g.,
by Gourgoulhon and Haensel (1993) and
Baumgarte et al. (1996a,b).

While all points on the $M(\rho_{\rm c})$ curve in Fig.
\ref{fig:str-Mrhoc} correspond to the configurations of hydrostatic
equilibrium, only the points which correspond to {\it stable
equilibrium} are of astrophysical interest. At a given total baryon
number,
hydrostatic equilibrium corresponds to the extremum of $M$ which
is a functional of matter density distribution $\rho$; this is
equivalent to the vanishing of the first-order variation of $M$,  due to
perturbation of $\rho$ denoted as $\delta\rho$. However, this does not
guarantee that the second-order variation, quadratic in
$\delta\rho$, is positive, which is necessary for the
hydrostatic stability.
It can  be shown (Harrison et al. 1965, Zeldovich and
Novikov 1971)
that stable equilibria correspond to the  segments of the
$M(\rho_{\rm c})$ curve for which ${\rm d}M/{\rm
d}\rho_{\rm c}>0$ ; this is the {\it static
stability criterion}.
Therefore, the configurations to the right of the maxima in
 Fig.\ \ref{fig:str-Mrhoc}  are unstable with respect to small radial
perturbations.\footnote[2]{Actual perturbations of neutron stars
are time-dependent, and can be decomposed into normal pulsation
modes. Assuming that pulsations do not move neutron-star matter
from its ground state (i.e., that the EOS for both static and
pulsating configurations is the same) one can show that the
configurations to the right of $M_{\rm max}$ in Fig.\
\ref{fig:str-Mrhoc} are indeed unstable with respect to the
fundamental mode of small radial pulsations (Harrison et al.
1966). The finite timescale of reactions between constituents of
neutron star matter which may be longer than the pulsation period
complicates the {\it dynamical} analysis of
 stability of neutron stars with respect to small
radial pulsations (Meltzner \& Thorne 1967, Chanmugan 1976,
Gourgoulhon  et al. 1995).
}
In view of this, $\rho_{\rm c,max}$ is the maximum density which
can be reached in the interior of stable static neutron stars.
 The value of the limiting density
$\rho_{\rm c,max}$ in static, cold neutron stars is even more
uncertain  than  the value of $M_{\rm max}$: it ranges from $2\times
10^{15}~{\rm g~cm^{-3}}$ (about $7\rho_0$) for the stiffest EOS to
 $4\times 10^{15}~{\rm g~cm^{-3}}$ (about $15\rho_0$) for the
softest  one. It should be stressed that even $7\rho_0$ is very
far beyond the limits within which our models of dense neutron star
matter can be considered as reliable and unambiguous.
\section{On the origin of $M_{\rm max}$}
\label{sect:Mmax.origin}
When explaining the very existence of $M_{\rm max}$,
two kinds of general physical arguments can  be used. They are
associated with  expected behavior of cold matter at ultrahigh
densities  and with General Relativity.

The arguments based on the high-density behavior of the EOS of
dense matter were proposed, for Newtonian stars,
 by Landau (1932). His brief paper
 presented a general physical argument why above some
$M_{\rm max}$ a large self-gravitating sphere of cold matter
cannot sustain itself against gravitational
collapse.\footnote[3]{Landau's paper was submitted for publication
before discovery of neutron.}
Harrison et al. (1965)  rephrased  the arguments in a way
applicable to neutron stars. However, it should be stressed that this way of
reasoning is based on the assumption  that in the high-density region in which the
stability is lost the EOS of dense matter is that of the ultra-relativistic free Fermi
gas. The counterexample to such a  behaviour was presented by Zeldovich (1962).

Irrespectively of the form of the EOS of dense matter, the upper bound on
$M$ is a consequence of the General Relativity. Consider the right-hand-side
of the OV equation of hydrostatic equilibrium, Eq.\ (\ref{str-P}). It describes
gravitational pull acting on a unit proper-volume of matter,
\begin{equation}
{\rm gravitational~pull~=} - {Gm\rho\over r^2}
\left(1 + {4\pi P\over m r^3}\right)
\left(1 + {P\over \rho c^2}\right)
\left(1 - {2Gm \over r c^2}\right)^{-1}~.
\label{struct-OV.rhs}
\end{equation}
The gravitational pull is given by a Newtonian-like term $-Gm\rho/r^2$, multiplied
by three relativistic factors.
 With increasing $M$, all three factors in brackets
amplify the pull compared to the Newtonian case, and the increase of $M$
with an increase of central pressure $P_{\rm c}$ becomes harder.
Mathematically,  the derivative ${\rm d}M/{\rm d}P_{\rm c}$ decreases with
growing $M$.

Let us illustrate this property with an {\it unphysical} case of incompressible
fluid of mass density $\rho_{\rm inc}={\rm constant}$.  The total gravitational mass
is then $(4\pi/3)\rho_{\rm inc}R^3$, and the pressure profile within the star
is determined analytically (solution was obtained by Karl Schwarzschild in 1916,
and
is 
described in Box 23.2 of Misner et al. 1973). The central pressure  $P_{\rm c}$
can be determined as a function of $r_{\rm g}/R$. As the mass increases,
$P_{\rm c}$ as well as $r_{\rm g}/R$ increase monotonically, too.
 For $P_{\rm c}\longrightarrow \infty$,  the radius tends to a finite
value $R_{M_{\rm max}}={9\over 8}r_{\rm g}$, and therefore the mass tends
to
\begin{equation}
M_{\rm max}^{\rm (inc)}
={4\pi\over 9}{R_{M_{\rm max}} c^2\over G}=
{4 c^3\over 3^{5/2}\pi^{1/2}G^{3/2}\rho_{\rm inc}^{1/2}}
=5.09~{\rm M}_\odot~\left(5\times 10^{14}~
{\rm g~cm^{-3}}\over \rho_{\rm inc}\right)^{1\over 2}
\label{struct-Mmax.inc}
\end{equation}
This limit is the General Relativity effect: there is no limit
to the mass of the incompressible-fluid
stars in Newtonian gravitation. If $M_{\rm max}$ exists
for an incompressible fluid,
then {\it a fortiori} it  should exist for {\it any} EOS of dense matter
with finite compressibility. However, the numerical value of $M_{\rm max}$
is mainly determined by the EOS  for $\rho\gtrsim 2\rho_0$ which
 is largely unknown.
\section{Upper bound on $M_{\rm max}$}\par
\label{subsect:struct-UBoundMmax}
In view of the uncertainty in the value of $M_{\rm max}$, it is of
interest to have a possibly firm  upper bound on $M_{\rm max}$.
 Let us assume that the EOS is known up to a certain density $\rho_{\rm
m}$. This reliably known segment of the EOS will be denoted by $P_<(\rho)$.
It spans the range $0<P\le P_{\rm u}\equiv P_<(\rho_{\rm u})$.
Having fixed  $P_<(\rho)$, we can treat $M_{\rm max}$ as a
{\it functional} of the EOS at $P>P_{\rm u}$; this {\it unknown}
EOS will be denoted as $P_>(\rho)$. The inverse function $\rho_>(P)$,
 determined for $P>P_{\rm u}$, does not need to be continuous. Let us  require
only that the $P_>(\rho)$ is subluminal
 and fulfills the
condition  of stability:
\begin{equation}
0< v_{\rm s}^2={{\rm d}P_>\over {\rm d}\rho}~\le c^2~,
\label{struct-EOS.stab.sublumin}
\end{equation}
where $v_{\rm s}$ is the speed of sound in the dense medium.
Our task is to find the maximum of the functional $M_{\rm max}
\left[P_>(\rho)\right]$ on a set $\lbrace P_>(\rho)\rbrace$ of
EOSs which satisfy conditions (\ref{struct-EOS.stab.sublumin}).
As it turns out, $M_{\rm max}$ is maximized by the so called
{\it causal-limit (CL) EOS},

\begin{equation}
P^{\rm CL}_>(\rho)=P_{\rm u}+(\rho-\rho_{\rm u})c^2~,
\label{struct-CL.EOS}
\end{equation}
which therefore yields the upper bound we are looking for, $M^{\rm
CL}_{\rm max}$. It is quite natural: the CL  EOS is the  stiffest
one  at $\rho>\rho_{\rm u}$. Numerical calculations show that for
$\rho_{\rm u}\lesssim 2\rho_0$ the effect of the outer layers with
$\rho<\rho_{\rm u}$ on the value of $M^{\rm CL}_{\rm max}$ is
negligibly small. Actually, had we used the {\it pure causal-limit
EOS} of the form $P=(\rho-\rho_{\rm s})c^2$ (where $\rho_{\rm s}$
is the surface density of a star  built of matter with such an
EOS), we would slightly increase (by less than one percent) the
value of $M_{\rm max}$ compared to that obtained for the CL EOS
with $\rho_{\rm u}=\rho_{\rm s}$~.\footnote[4]{ For a pure CL EOS
one gets an exact result $M_{\rm max}=3.0\; (5\times 10^{14}~{\rm
g~cm^{-3}}/\rho_{\rm s})^{1/2}\;{\rm M}_\odot$~.}
This can be easily explained, because $P_<(\rho)$ is  not maximally
stiff. All in all, for $\rho_{\rm u}\lesssim 2\rho_0$ one gets
(Rhoades and Ruffini 1974, Hartle 1978, Kalogera and Baym 1996, Glendenning 1999)
\begin{equation}
v_{\rm s}\le c~~~
\Longrightarrow~~~
M_{\rm max}\le M^{\rm CL}_{\rm max}=
3.0\left(
{5\times
10^{14}~{\rm g~cm^{-3}}\over \rho_{\rm u}}
\right)^{{1\over 2}}~{\rm M}_\odot~.
\label{str-CLMmax}
\end{equation}
The maximum allowable mass for any EOS($\rho>\rho_{\rm u}$) with
subluminal sound velocity is lower than the above upper bound. As
the inequality (\ref{str-CLMmax}) is widely accepted, it seems to
be safe to say that actual $M_{\rm max}$ of static neutron stars
built of baryonic matter is below $3~{\rm
M}_\odot$.\footnote[5]{Notice that we assume that neutron star has
an outer layer of density $\rho<\rho_{\rm u}$, composed of normal
cold dense matter. For some very exotic hypothetical models of
compact stars with superdense surface built, e.g.,  of a
self-bound Q-matter, the limit  of $3~{\rm M}_\odot$ does not
apply (Sect.\ \ref{sect:stars.abnormal}).}
A rapid rotation
increases the value of $M_{\rm max}^{\rm CL}$, as discussed
in Sect.\ \ref{sect:Mmax.rot}.
\section{Phase transitions in dense matter, third family
of compact stars,   and maximum mass}
\label{sect:struct-phase.tr}
The existence of exotic
phases of dense matter in neutron star cores is a subject of
debate. Unfortunately, since the
 physics of dense matter at $\rho\gtrsim 2\rho_0$ is uncertain,  the
problem cannot be solved within the
present many-body theory.  Therefore one is forced to
admit that the question: {\it is a specific exotic phase of dense matter
present in a neutron star core?}
 can only be answered by unambiguous identification of
signatures of  this phase in neutron
star observations.

\begin{figure}
\begin{center}
\includegraphics[width=10cm]{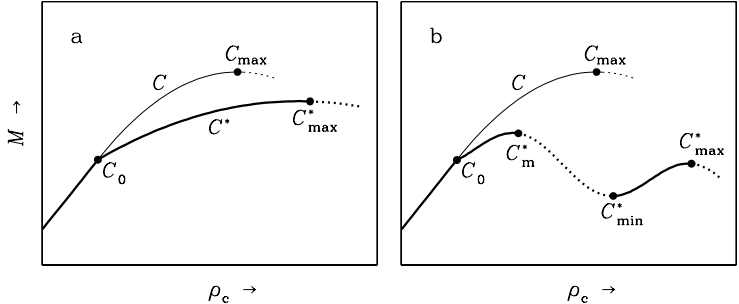}
\caption{
Mass -- central-density relations for
an EOS, for which a phase transition
at a pressure $P_0$ implies softening
with no density jump 
(right panel of Fig. 3). 
 Configuration ${\cal C}_0$ is the last one
composed exclusively of a lower-density phase, its  central
pressure $P_{\rm c}=P_0$.
 Thin lines
represent configurations calculated using the  EOS without a phase
transition. Thick lines show  configurations calculated
for the  EOS with a phase transition. Dotted segments correspond to
configurations which are unstable with respect to small radial
perturbations.  Left panel (a): moderate softening
of the EOS.  Right panel (b):
 strong softening of the EOS,
implying appearance
of an unstable branch between ${\cal C}^*_{\rm m}$ and ${\cal
C}^*_{\rm min}$ (thick dotted line),
and a {\it separate branch of superdense stars} between
${\cal C}^*_{\rm min}$ and ${\cal C}^*_{\rm max}$. }
\label{fig:struct-Mrhoc.p.t.1}
\end{center}
\end{figure}
\begin{figure}
\begin{center}
\includegraphics[width=10cm]{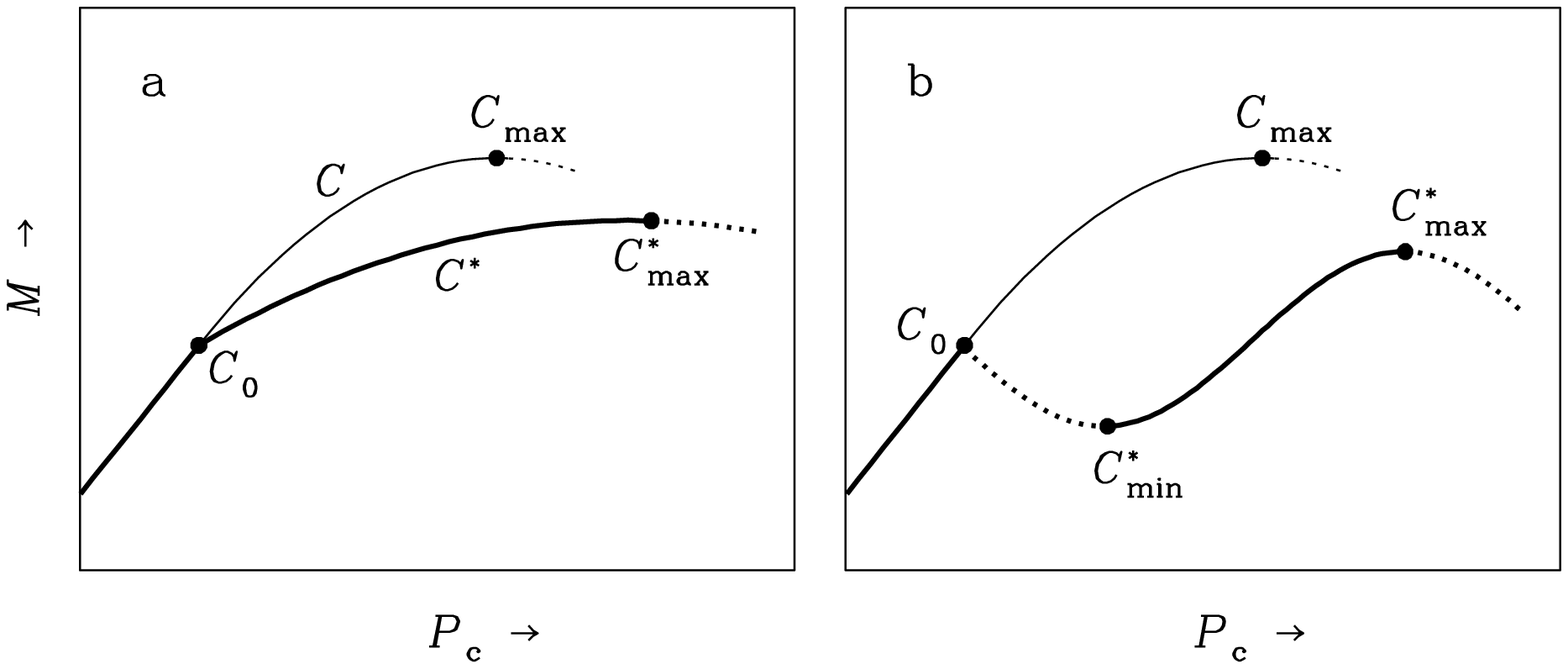}
\caption{
Mass -- central-pressure relations for
an EOS  containing  a phase transition  with  a
density jump
(left panel of Fig. 3).
 Configuration ${\cal C}_0$ is the last one composed
exclusively of a lower-density phase, with $P_{\rm c}=P_0$.
Notations are the same as in Fig.\ \ref{fig:struct-Mrhoc.p.t.1}.
Panel (a):  moderate softening of the EOS.
 Panel (b):
 strong softening of the EOS, implying appearance of an unstable branch
between ${\cal C}_{0}$ and ${\cal C}^*_{\rm min}$ (thick dotted line),
and a separate
branch of {\it superdense stars} between ${\cal C}^*_{\rm min}$ and
${\cal C}^*_{\rm max}$.
For further explanation see the text.}
 \label{fig:struct-MPc.p.t.2}
\end{center}
\end{figure}

In  what follows, we will distinguish the case of a moderate
and  a strong softening of an EOS by a phase transition.
We will refer to a phase transition
as {\it moderate}, if it does not lead to the appearance of an unstable
segment in the $M-\rho_{\rm c}$  curve. In
the opposite case of a {\it strong softening} of the EOS, the phase
transition will imply the existence of an unstable segment of the
 $M-\rho_{\rm c}$ curve, which will separate
a family of lower-density configurations from a distinct family of
superdense compact objects.
Let us start with the case of the moderate softening of the EOS,  without
density jump (right panel of Fig.\ \ref{fig:EOSpt}).
 Then $\rho(P)$ is continuous,
$\lambda\equiv \rho_{\rm h}/\rho_{\rm l}=1$,
but the adiabatic index $\gamma=(n_{\rm b}/P){\rm
d}P/{\rm d}n_{\rm b}$ suffers a jump (a drop) at
the appearance of a new phase at $P=P_0$.
The $M-\rho_{\rm c}$ curve is displayed in
Fig.\ \ref{fig:struct-Mrhoc.p.t.1}a.
Such a
situation corresponds to hyperonization, and also to transitions
to a mixed phase (baryons and pion or  kaon condensate, baryons
and deconfined quark matter). In the latter case, the mixed phase
exists up to some limiting density at which a transition from a
mixed phase to a pure higher-density phase takes place; it is is
accompanied by a second discontinuous change (actually,  an
increase) in $\gamma$.

Consider now a phase transition which implies a
strong softening of the EOS without  a density jump.
This is the extreme case of a very strong decrease of $\gamma$
after the phase transition discussed above, followed by a hardening of
the EOS at still higher densities. In this extreme
situation the $M-\rho_{\rm c}$  curve has two
stable branches of static equilibrium configurations,
Fig.\ \ref{fig:struct-Mrhoc.p.t.1}b, corresponding to two
distinct families of neutron stars.
The first (lower-density or normal)
family
is continuously connected with low-mass
neutron stars and terminates at ${\cal C}^*_{\rm m}$.
The second family contains  ``compact neutron
stars'', and will be referred to as the ``higher-density family''
or ``superdense  family''. The configurations belonging to the
 superdense  family have central densities
 $\rho^*_{\rm min}<\rho^*_{\rm c}<\rho^*_{\rm max}$
 and masses $M^*_{\rm min}<M^*<M^*_{\rm max}$. They are more compact
 and tightly bound than those
containing the same number of baryons
  and  belonging to the  normal branch.

 The appearance of a distinct family of superdense stars deserves
 a   general remark. Superdense stars form actually the  {\it third
 family of compact stars}  built of degenerate matter,
 the remaining two being the well known white dwarfs and
  ``lower-density'' neutron stars.

Two examples of the strong softening of the EOS without  density
jumps are given by Glendenning and Kettner (2000) and
Schaeffner-Bielich et al. (2002). In the model of Glendenning and
Kettner (2000) the softening of the EOS results from the
appearance of a quark-baryon mixed phase and takes  place
for a substantial quark fraction. At higher densities,
their EOS stiffens   when the
mixed phase is replaced by a pure quark phase. In the model
of Schaeffner-Bielich et al. (2000) the softening is due to a
copious appearance of hyperons. It is  obtained for a
specific model of hyperon-hyperon
interaction. It is followed by a substantial stiffening of the
EOS at higher densities. Similarly to  the first example, it
leads to the appearance of the  second stable ``higher-density branch''
of the $M-\rho_{\rm c}$  curve; the models belonging to this branch
are mostly composed of hyperons. In both examples, the maximum
mass of the ``higher-density branch'' is lower than that of
the ``lower-density'' family, $M^*_{\rm max}<M_{\rm max}$.

 At the same baryon number, the energy $M^* c^2$ of the high-density
 configuration is lower.
Let $A^*_{\rm min}$ and $A^*_{\rm max}$  denote the
limiting baryon numbers of stable configurations on the
 higher-density branch. Configurations with extremal mass and extremal baryon number
coincide (see, e.g., Harrison et al. 1965).
 For $A^*_{\rm min}<A<A^*_{\rm max}$,  the true stable configurations are those
 belonging to the higher density branch, while the
configurations  consisting of
 the same number of baryons and lying on the lower-density branch
 are {\it metastable}.

As a third type of phase transition in dense matter, we consider
a transition with a moderate softening associated with a density jump
in the EOS. This  jump occurs
 at a pressure $P_{\rm 0}$, at which a pure lower-density l-phase of density
 $\rho_{\rm l}$ coexists with a pure higher-density {\bf h-phase}
  of density $\rho_{\rm
 h}$, so that the relative density jump
 is $\lambda=\rho_{\rm h}/\rho_{\rm l}>1$ (left panel of
 Fig.\ \ref{fig:EOSpt}).
 In view of
 the discontinuity of $\rho_{\rm c}$ at $P_0$, the equilibrium
 configurations have  to be parametrized by the central pressure
 $P_{\rm c}$.

One has to stress a particular role played by
the quantity
\begin{equation}
\lambda_{\rm crit}\equiv {3\over 2}
\left(1 +{P_0\over \rho_{\rm l}c^2}
\right)~.
\label{lambda.crit}
\end{equation}
A moderate
softening associated with a first order phase transition corresponds
to $\lambda<\lambda_{\rm crit}$.

Finally, we will consider the strong softening of the EOS
associated with a first order  phase transition with $\lambda
=\rho_{\rm h}/\rho_{\rm l}>\lambda_{\rm crit}$. In this case, the
appearance of a small core of  h-phase destabilizes the neutron
star. Specifically, for static configurations ${\cal C}^*$ with a
small h-core  and $\lambda>\lambda_{\rm crit}$ one has ${\rm
d}M/{\rm d}\rho_{\rm c}<0$. Such configurations are therefore {\it
unstable} with respect to small radial perturbations and collapse
into {\it stable}   configurations with a large  h-core. The {\it
instability condition} $\lambda>{3\over 2}(1+P_0/\rho_{\rm l}
c^2)$ was first derived by Seidov (1971), using the static energy
method. Ten years later this condition was rediscovered by
Kaempfer (1981) who studied the necessary condition for the onset
of the neutron-star collapse initiated by the phase transition at
its center. It is worth to mention that the Newtonian version of
this criterion ($\lambda>{3\over 2}$) was first derived by
Lighthill (1950, see also Ramsey 1950) in the context of stability
of two-phase planets. The relativistic effects stabilize neutron
stars with small h-cores by $\Delta\lambda_{\rm crit}=
\lambda_{\rm crit}-{3\over 2}={3\over 2}P_0/\rho_{\rm l} c^2$,
which can be as high as $\sim 0.2$. Static stable equilibrium
configurations of neutron stars split into two families,
visualized in  Fig. \ref{fig:struct-MPc.p.t.2}b. The superdense
branch ${\cal C}^*_{\rm min}{\cal C}^*_{\rm max}$ forms a {\it
third family of compact stars}, apart from the white dwarfs and
lower-density neutron stars. It should be stressed, that in
contrast to the instability implied by a second-order phase
transition due to the hyperonization or appearance of a mixed
quark-baryon  phase (Sect.\ \ref{sect:EOScore}), {\bf a}
 typical situation for
 $\lambda>\lambda_{\rm crit}$ corresponds to
 $M_{\rm max}<M^*_{\rm max}$ and  $A_{\rm max}<A^*_{\rm max}$ (see, e.g.,
 Brown and Weise 1976, Haensel and
 Pr{\'o}szy{\'n}ski 1982, and Migdal et al. 1990).

\section{Rotating neutron stars}
\label{sect:NSrot}
We  limit ourselves to a stationary rigid rotation.
In this case   the angular frequency  $\Omega$ of rotation of
a matter element, measured by an observer at infinity,
is constant.  A rotating configuration has axial symmetry with respect
to the rotation axis. A stationary rotation of stellar bodies in General Relativity
has been studied by many authors;  the present state of this field is reviewed
in a {\it Living Review} by Stergioulas (2001).

 Rotating  configurations form a two-parameter
family, and can be labeled, e.g., by the values of $\rho_{\rm c}$ and $\Omega$.
 These configurations are flattened and their equatorial radius,
 $R_{\rm eq}$,  is larger than the polar
radius, $R_{\rm pol}$. Configurations ${\cal C}(\rho_{\rm c},\Omega)$ cover a region
in the $M-R_{\rm eq}$ plane. A curve $M(R_{\rm eq},\Omega)$ for a fixed $\Omega$ is
limited on the high- and  low-density sides. On the high-density side, the curves
are limited by the condition of  stability  with respect to  small axi-symmetric
perturbations.  In order to check whether a configuration ${\cal C}(\rho_{\rm
c1},\Omega_1)$ with angular momentum $J_1$ is stable, one has to consider a family of
configurations ${\cal C}(\rho_{\rm c},J_1)$ with fixed  angular momentum $J=J_1$  in
the neighborhood of ${\cal C}(\rho_{\rm c1},\Omega_1)$.
Configuration ${\cal C}(\rho_{\rm c1},\Omega_1)$ is stable if
\begin{equation}
\left[
\left( 
{\partial M\over \partial \rho_{\rm c}} 
\right)_{J=J_1}
\right]_{\rho_{\rm c}=\rho_{\rm c1}}>0~, 
\label{struct-stab.cond.rot}
\end{equation}
%
where the derivative is calculated along the $\lbrace{\cal C}(\rho_{\rm
c},J_1)\rbrace$ family.  A line determined by
 $\left(\partial M/\partial \rho_{\rm c}\right)_J=0$
separates the configurations stable with respect to small
axisymmetric perturbations from the unstable ones.\footnote[6]{In
the static limit, $J_1=0$, we recover the condition of stability
with respect to radial perturbations (which are a special case of
axisymmetric perturbations), ${\rm d}M/{\rm d}\rho_{\rm c}> 0$
(see Sect.\ \ref{sect:stability}).}

The low-density boundary is determined by  the condition of stability with respect to
the mass-shedding from the equator. A necessary condition for the existence of a
stationary rotating configuration is the following: the equatorial velocity of an
element of stellar matter  has to be smaller than the
velocity of a test particle moving on a circular orbit of radius $R_{\rm eq}$ in the
equatorial plane, the so-called Keplerian velocity $U_{\rm K}$. The
Keplerian velocity corresponds to the {\it Keplerian angular frequency} $\Omega_{\rm
K}=U_{\rm K}/R_{\rm eq}$, called also the {\it mass-shedding} angular frequency.

In order to illustrate the effect of rotation on the structure of {\it observed
pulsars}, consider the radio pulsar  PSR B1937+21
with shortest observed period effect of rotation   $P^{\rm
obs}_{\rm min}=1.56~$ms (Backer et al. 1982);  
the corresponding angular frequency is 
$\Omega^{\rm obs}_{\rm max}=641~$Hz. 
The effect of rotation on  the maximum-mass configuration is
very small. For $M\simeq M_{\rm max}$ the period of 1.56 ms corresponds to the {\it
slow rotation regime} ($\Omega\ll \Omega_{\rm K}$), and  therefore $M^{\rm
1.56ms}_{\rm max}-M_{\rm max}^{\rm stat}$ is quadratic in the small parameter
$\overline{\Omega}=\Omega/\sqrt{GM/R^3}$. Consider for example the specific case of
the SLy EOS (Douchin and Haensel 2001). Then
$\overline{\Omega}^2=(\overline{\Omega}_{\rm max}^{\rm obs})^2 \simeq 0.06$. As
will be shown  in Sect.\ \ref{sect:Mmax.rot}, the highest rotation
 frequency  allowed by the condition of stationarity implies the increase
 of  $M_{\rm max}$ by  some 20\%. Therefore, the fractional increase of
 $M_{\rm max}$ connected with rotation at $P=1.56$~ms is $0.2
 (\overline{\Omega}_{\rm max}^{\rm obs})^2
 \simeq 2\%$, which agrees very well with the exact numerical result
 (Haensel \& Douchin 2001). The effect will be smaller
for a softer  EOS (e.g., FPS EOS) and larger for a stiffer
  EOS (e.g., APR EOS).
\section{Maximum mass  of rotating neutron stars}
\label{sect:Mmax.rot}
Rotation increases $M_{\rm max}$ because the centrifugal forces
oppose the gravity. For $\Omega\ll \Omega_{\rm K}$, this increase
is quadratic in $\Omega/\Omega_{\rm K}$ and therefore very small.
More generally, $M_{\rm max}(\Omega)-M_{\rm max}^{\rm stat}$ is an
even function of $\Omega$, because it does not depend on the
orientation of the spin axis.\footnote[7]{We remind that our stars
are built of an ideal fluid and the effects of the magnetic fields
are neglected.}

Within the entire set of stable stationary  configurations
$\lbrace{\cal C}(\rho_{\rm c},\Omega)\rbrace$  one can
find two important
extremal configurations: with the maximum mass $M^{\rm rot}_{\rm max}$ and
with the minimum period $P_{\rm min}$.
These two configurations do not coincide
but  are very close to each other. Depending on the EOS, the {\it maximally
rotating configuration} with the rotation period $P_{\rm min}$
can have the central
density higher or lower than the maximum-mass configuration, but the difference
is at most a few percent (Cook et al. 1994). As a rule, the mass of the maximally
rotating configuration is lower than $M_{\rm max}^{\rm rot}$ by less than
one  percent (Cook et al. 1994).

It is useful to note that for realistic baryonic, subluminal
 ($v_{\rm s}\le c$) EOSs,
$M_{\rm max}^{\rm rot}$ is (to a very good approximation,
typically within 3\%) proportional
to the maximum mass of non-rotating  configurations (Lasota et al. 1996)
\begin{equation}
M_{\rm max}^{\rm rot}\simeq 1.18 M_{\rm max}^{\rm stat}~.
\label{struct-Mmax.rot.stat}
\end{equation}
However, the above formula is not valid for the EOSs of the
form $P/c^2=a(\rho-\rho_{\rm s})$, and in particular, for strange
quark stars; their case will be considered in
Sect.\  \ref{sect:strange}.

In Sect.\ \ref{subsect:struct-UBoundMmax} we derived an absolute upper
bound on the static neutron-star mass based on the knowledge of the
EOS at $\rho<\rho_{\rm u}\sim 2\rho_0$ under the constraint of
$v_{\rm s}\le c$. Let
$M^{\rm CL,stat}_{\rm max}$
 be the upper bound for non-rotating  neutron stars.
Rotation will increase the
upper bound, $M^{\rm CL}_{\rm max}(\Omega)>
M^{\rm CL,stat}_{\rm max}$. The upper bound for rotating stars,
$M^{\rm CL,rot}_{\rm max}$,
is   obtained for the same causality-limit EOS as for the
non-rotating models;  it is reached at   $\Omega$ very close
to $\Omega_{\rm max}$. Its precise value
 was obtained  by Koranda et al. (1997)
\begin{equation}
v_{\rm s}\le c~:~~~
M_{\rm max}^{\rm rot}\le
M^{\rm CL,rot}_{\rm max}=3.89~{\rm M}_\odot
\left(
{\rho_{\rm u}\over
5\times 10^{14}~{\rm g~cm^{-3}}}
\right)^{-{1\over 2}}~.
\label{struct-Mmax.rot}
\end{equation}
For a given $\rho_{\rm u}$, it is 30\% larger  than the static upper bound,
$M^{\rm CL,stat}_{\rm max}$.

The rapidly rotating configurations, considered in Sects.\ \ref{sect:NSrot}--\ref{sect:Mmax.rot},
are  stable with respect to the axisymmetric perturbations and mass shedding. However, they
can be susceptible to various {\it secular instabilities}, reviewed by Stergioulas (2001).
\section{Strange quark stars}
\label{sect:strange}
 By strange stars we will mean the  compact objects
built entirely, or predominantly, of self-bound strange quark matter
(SQM, Sect.\ \ref{sect:EOSstrange}).
The possibility of the existence of self-bound strange quark stars
built {\it entirely} of SQM,
 was contemplated
by  Witten (1984), who  considered a simplified
model of SQM, composed of massless u, d, and s quarks,
with the EOS of the form $P={1\over 3}(\rho c^2 - B)$ (see also
 Brecher and Caporaso 1976). Witten  showed
that for the bag constant $B\approx 60~{\rm MeV~fm^{-3}}$
 close to the value  needed to reproduce
experimental masses of baryons within the MIT Bag Model,
 the parameters of the maximum-mass configuration
 for strange stars are similar to those  for realistic
neutron
stars  built of baryonic matter. The first detailed models of strange
stars, based on a more realistic EOS of SQM, taking into
account strange quark mass and the lowest-order QCD interactions, were
constructed by Haensel et al. (1986) and Alcock et al. (1986). These
authors discussed the  basic physical
properties of strange stars,  and their astrophysical
manifestations.

Further development of physics and astrophysics of strange stars
 focused on the refinement of the EOS of SQM
(particularly, beyond the  MIT Bag
Model), and on specific
 properties of strange stars, such as neutrino
emissivity, rotational properties, superfluidity, pulsations,
electromagnetic radiation,
and cooling. Physics and astrophysics of strange stars is
reviewed
 by Glendenning (1999), Weber (1999),
 and
Madsen (1999).

The surface density  of bare strange stars is equal to
that of SQM at zero pressure, $\rho_{\rm s}$. It is therefore some
fourteen orders of magnitude larger than the surface density of
 normal neutron stars.

\begin{figure}
\begin{center}
\includegraphics[width=8cm]{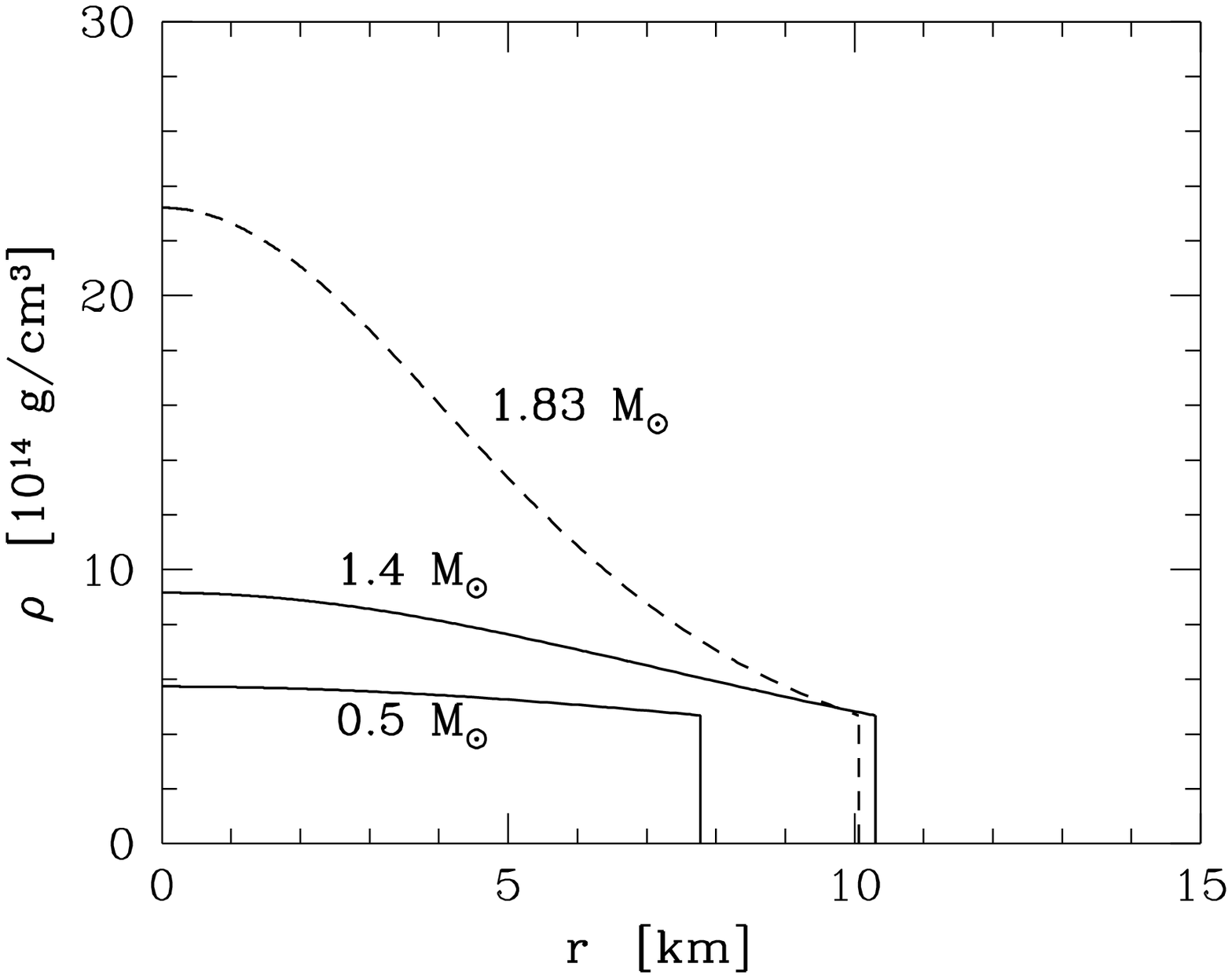}
\caption{
Mass density  $\rho$ versus radial
coordinate $r$ for three bare strange stars  of different
masses, calculated for the SQM1 EOS of strange quark matter.
 The highest of
three masses is the maximum allowable mass for this EOS. }
\label{fig:rho.prof.SS}
\end{center}
\end{figure}

In what follows, we will illustrate the generic properties of strange
 stars assuming  SQM1  EOS of SQM of Zdunik et al. (2001).
\begin{figure}
\begin{center}
\includegraphics[width=11cm]{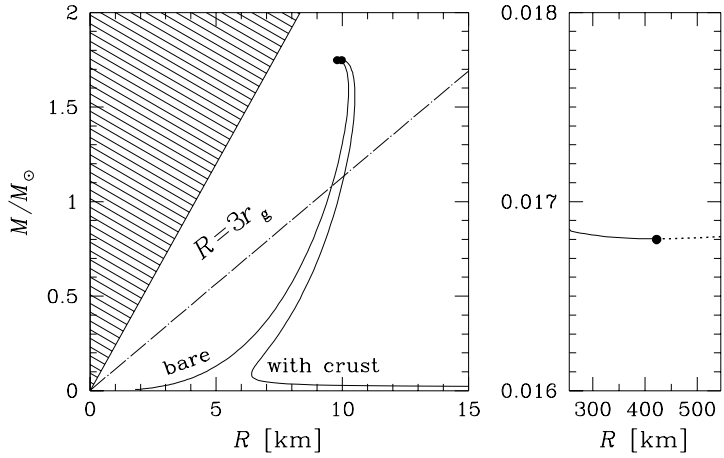}
\caption{
Mass-radius relation for bare strange stars and for strange stars
with normal crust calculated for the SQM1 EOS.
Extremal (maximum and minimum mass) configurations
are indicated by filled circles.
 Strange stars
with crust have maximum crust mass and thickness:
density at the crust bottom
is  equal to the
neutron-drip one, $4.3\times 10^{11}~\mdens$. The dotted segment represents
configurations unstable with respect to small radial
perturbations. The hatched area corresponds to the region of the $M-R$ plane
prohibited  by General Relativity and by condition $v_{\rm s}\le c$.
Long dash-dot line gives the radius of the marginally stable circular
orbit around a strange star 
(it is the radius of the innermost stable orbit of a particle 
orbiting in the equatorial plane of a non-rotating star). 
 The vicinity of the minimum-mass configuration
for strange stars with
the crust is shown in the right panel.
}
\label{fig:strange-MR.SS}
\end{center}
\end{figure}
 The $M-R$ curve for bare strange stars
   is shown in
 Fig.\ \ref{fig:strange-MR.SS}. For strange stars with
 $M>1~{\rm M}_\odot$ the radius changes very little with $M$,
 $R\simeq 9-11$~km. It  is
 quite similar to that obtained for neutron stars with a moderately
 stiff EOS. However, at lower masses, the
 radius of bare strange stars behaves in a completely different
 way. Namely, it decreases
 monotonically with decreasing $M$, and
 $R\propto M^{1\over 3}$
for $M\lesssim 0.3~{\rm
 M}_\odot$.
  Such
 a  behavior of low-mass bare strange stars can easily be explained
 within the bag model.
 Gravitational pull decreases rapidly with decreasing $M$,
and can be neglected
 at
 $M\lesssim 0.3~{\rm M}_\odot$
compared to the
pressure of the normal vacuum on the  volume ``filled'' by the QCD
vacuum: this pressure
 confines  SQM to a  sphere of radius $R$. Due to very
high incompressibility of strange matter,  the density
within a low-mass strange star is nearly constant and close to $\rho_{\rm s}$
(see Fig.\ \ref{fig:rho.prof.SS}). On the other hand, the low-mass
strange stars can be described in the Newtonian theory, which
gives  $M\simeq {4\pi\over 3}\rho_{\rm s}R^3$ and
$R\propto M^{1\over 3}$.

The internal structure of bare strange stars is very different from that of
neutron stars. First of all, their surface density is huge,  $\rho_{\rm s}
\sim 10^{15}~\mdens$. The density profile in such a   star
 is very flat (Fig.\ \ref{fig:rho.prof.SS}).
Even at the maximum mass, i.e., under the maximum gravitational
compression, the central density is only five times higher than
the surface one; this is to be compared  with fourteen orders of magnitude
center-to-surface density
difference for neutron stars! The   density
difference decreases rapidly with decreasing  $M$. For a $1.4~{\rm
M}_\odot$ bare strange star the central density is only forty percent
higher than the surface one. For low-mass bare strange stars
of $M\lesssim 0.3~{\rm M}_\odot$,  the center-to-surface density difference is
negligibly small.

\subsection{Maximum mass of strange quark stars}
\label{subsect:strange-scaling}
As mentioned before, the EOS
of SQM is  accurately   determined by two
parameters, $a$ and $\rho_{\rm s}$
(Sect.\ \ref{sect:EOSstrange}).
Consider bare strange quark stars.
Using the linear
representation  of
the EOS
 and passing to
dimensionless variables  ${\widetilde P},~{ \widetilde \rho},
~{\widetilde r}$,
%
\begin{equation}
{\widetilde P}={P\over \rho_{\rm s}c^2}~,
~~~~{\widetilde\rho}={\rho\over \rho_{\rm s}}~,
~~~~{\widetilde r}=r\cdot \sqrt{{G\rho_{\rm s}\over c^2}}~,
\label{strange-Prho.r.dimless}
\end{equation}
one can rewrite the equations of hydrostatic equilibrium of
strange stars in a dimensionless form,
At a fixed $a$, the equilibrium configuration can be
obtained from a universal dimensionless solution of transformed
Eqs.\ (\ref{str-P})--(\ref{str-m})
 by returning to ordinary variables $r,~P,~\rho$.
 This property of Eqs.\ (\ref{str-P})--(\ref{str-m})
 implies useful scaling properties
relating  a solution which corresponds  to an  EOS $P=a(\rho
-\rho^\prime_{\rm s})c^2$  to the  solution for the EOS
$P=a(\rho-\rho_{\rm s})c^2$. In
particular, the  maximum mass
 for different values $\rho_{\rm s}$ and
$\rho^\prime_{\rm s}$ are related   by the {\it scaling relation}

\begin{equation}
M^\prime_{\rm max}=M_{\rm max}\cdot \left({\rho_{\rm s}\over
\rho^\prime_{\rm s}}\right)^{1\over 2}~,
\label{strange.Mmax.scal}
\end{equation}

For a strange matter EOS based on the MIT Bag Model, the  scaling
relation, Eq.\ (\ref{strange.Mmax.scal}),
 turns out to be precise within better than one percent
 (Haensel et al. 1986).
Similar precision is obtained for other models of SQM
(Gondek-Rosi{\'n}ska et al. 2001).

Scaling properties  become particularly simple
(and actually exact) for
 the MIT Bag Model EOS with
non-interacting massless quarks, i.e., for $a={1\over 3}$ and
$\rho_{\rm s}=4B/c^2$. This EOS will be  referred
to as SQM0. The formula for
$M_{\rm max}$ for the SQM0 EOS was  derived by
Witten (1984),
\begin{equation}
M_{\rm max}={1.96\over \sqrt{B_{60}}}~{\rm M}_\odot~,
\label{strange.Mmax.SQM0}
\end{equation}
where $B_{60}=B/(60~{\rm MeV~fm^{-3}})$.

The scaling formula  enables  one to explain why the parameters of
massive strange stars  ($M\sim 1-2~{\rm M}_\odot$) are so similar
to what we expect for normal neutron stars. Consider first the
simplest SQM0 model. For the
hypothesis of strange matter to be correct, energy per unit baryon
number in SQM should be less than $930.4~$MeV 
which is the minimum energy per nucleon in normal baryonic 
matter at zero pressure and temperature; this value is reached for the 
$^{56}{\rm Fe}$ crystal. 
This condition implies
$B_{\rm 60}<1.525$.
 On the other hand, a consistent model of SQM
should not lead to a spontaneous fusion of neutrons into droplets
of u, d quarks, which would eventually transform into droplets of
SQM -- ``strangelets''. This in turn means that the energy per unit
baryon number in the u, d matter with $2n_{\rm d}=n_{\rm u}$  should be above
939.6 MeV, which implies $B_{60}>0.982$.
 All in all, both
constraints lead to $1.6~{\rm M}_\odot <M_{\rm max}< 2.0~{\rm
M}_\odot$, which reminds us a typical maximum mass range
calculated using  medium-stiff EOSs of baryon matter. An allowance
for the finite mass of the strange quark and for  the QCD
interactions  within the MIT Bag Model  can increase  maximum
masses of strange stars due to a simultaneous decrease of the
allowed values of $B$ (this results from the condition $E_{\rm
SQM}<930.4$~MeV, see Zdunik et al. 2000).

A strange star can be covered by a solid crust of matter with
density lower than the neutron drip point (outer neutron-star crust).
The nuclei of the crust do not fuse with the SQM core because
they are  separated
from the SQM by a  repulsive Coulomb barrier.
However, since the mass of the crust is lower than  $10^{-4}~{\rm M}_\odot$,
its effect on $M_{\rm max}$ of strange stars is negligible.
\subsection{Maximum mass of rotating strange quark stars}
\label{sect:MmaxSSrot}
Rotation increases the maximum mass of strange quark stars
stronger   than $M_{\rm max}$ of neutron stars. This difference
 results from the different matter distribution within strange
quark stars: the density profile is relatively flat and the surface density
is huge. This feature increases  the effect of centrifugal forces
on the stellar structure. For simplest SQM0 EOS
 one gets the exact result
(Gourgoulhon et al. 1999)
\begin{equation}
{\rm SQM0~EOS}~:~~~~~~~~~~M^{\rm rot}_{\rm max}=
1.44 M^{\rm stat}_{\rm max}~.
\label{MmaxSSrot}
\end{equation}
In the case of more realistic EOSs of SQM  relation (\ref{MmaxSSrot})
holds only approximatively. Still, on may say that the increase
 of $M_{\rm max}$ for strange quark stars due to rapid rotation
 (about 40\%) is twice that for neutron stars (20\%).
\section{Maximum mass of  Q-stars}
\label{sect:stars.abnormal}
The models of hypothetical Q-stars were constructed by Bahcall et al.
(1990). As  mentioned in Sect.\ \ref{sect:EOSabnormal},  two
basic parameters of the model are: the
energy density  $U_0$ of the scalar
field inside the Q-matter  and
the coupling strength $\alpha_v$ of the vector field to
nucleon. The results for
Q-star parameters show a simple scaling with $U_0$ provided
 the  dimensionless
parameter $\zeta=\alpha_v U^{1/2}_0\pi/\sqrt{3}$ (Sect.\ \ref{sect:EOSabnormal})
is kept constant.
Then, the scaling  of $M_{\rm max}$
 with $U_0$  is the same as that with respect to
$\rho_{\rm s}$ for strange quark stars, Eq.\ (\ref{strange.Mmax.SQM0}),
 but with different numerical coefficients. For
$0\le\zeta\le 16$ and $U_0=13.0~{\rm MeV/fm^3}$,  one gets an
astonishingly  high maximum allowable mass, $4.0\le M_{\rm max}\le
8.3~{\rm M}_\odot$. This stems from a  low value of
Q-matter density $\rho_{\rm s}$ at zero pressure: remember that
 nucleons are nearly massless in the Q-matter.  Consider the simplest case
of $\zeta=0$. Then,  the EOS for the Q-matter coincides with
that for strange matter of massless, non-interacting quarks.
 However, the standard value used  by Bahcall et al. (1990) while  constructing
the families of the Q-stars  is $U_0=13.0~{\rm
MeV~fm^{-3}}$, which  corresponds to $\rho_{\rm s}=1.0\times
10^{14}~\mdens~$, 
only one third of the normal nuclear density! 
The maximum allowable mass can then be
calculated using Eq.\ (\ref{strange.Mmax.SQM0}), and
is $4.0~{\rm
M}_\odot$, in agreement with Figs.\ 4 and 8 of Bahcall et al. (1990).
With increasing $\zeta$, the EOS of the Q-matter becomes stiffer. In
the limiting case $\zeta=16$  considered by  Bahcall et al. (1990)
(at the same value of $U_0=13.0~{\rm MeV}~\bdens$)  they get
$\rho_{\rm s}=5.5\times 10^{13}~\mdens$ and $M_{\rm max}=
8.2~{\rm
M}_\odot$. This is not  surprising: for $a=1$ ($v_{\rm s}=c$)
we can use a pure-causal limit formula of Sect.\ \ref{subsect:struct-UBoundMmax},
getting  an estimate  of the maximum mass
$\approx 9.0~{\rm M}_\odot$. This estimate
 reproduces within 8\% the exact  value of $M_{\rm max}$,
 obtained by  Bahcall et al. (1990). Rapid rotation could further
 increase the maximum mass of Q-stars by some 30\%, to well above
 10 ${\rm M}_\odot$.

Summarizing, while the arguments for the existence of Q-matter
stem from sophisticated supersymmetric extensions of
field-theoretic models of dense nucleon matter, the practical
reasons for shockingly high $M_{\rm max}$ for Q-stars are very
simple. Namely, the predicted density of self-bound Q-matter at
zero pressure is three  to five times lower (!) than the normal
nuclear density, which results from a strong reduction of
effective nucleon masses in this hypothetical  state of matter.

Additional complication
 is that the typical average density of Q-stars is significantly
lower than the nuclear density. Therefore, a conversion of a neutron
star into a Q-star should be accompanied by a significant
inflation of stellar size, which can be obtained only at the
expense of a gigantic work done against the gravitational pull. This
should be  a very peculiar type of a transformation, in which a compact star
becomes less dense, but  more bound,  because nearly all
rest mass of nucleons  annihilated in the
phase transition.  Finally, while the existence of supersymmetric
(Q) ground-state of matter  may be not in conflict with the terrestrial
nuclear physics, reaching this state during stellar  evolution may
be virtually impossible due to huge energy barrier separating the
Q-matter
from the normal state of dense matter.
\section{Confronting theory with observations}
\label{sect:observ}
Some  neutron stars belong to
 binary stellar systems. In these cases, one can try
to measure their masses by analyzing orbital motion of the binary.
Up to now more than one hundred  such binary systems
containing neutron stars have been discovered which are quite
different in nature and  observational appearance. In what follows
we briefly review evaluations of neutron star masses based on the
analysis of binaries containing neutron stars.

\begin{figure}
\begin{center}
\includegraphics[width=10cm]{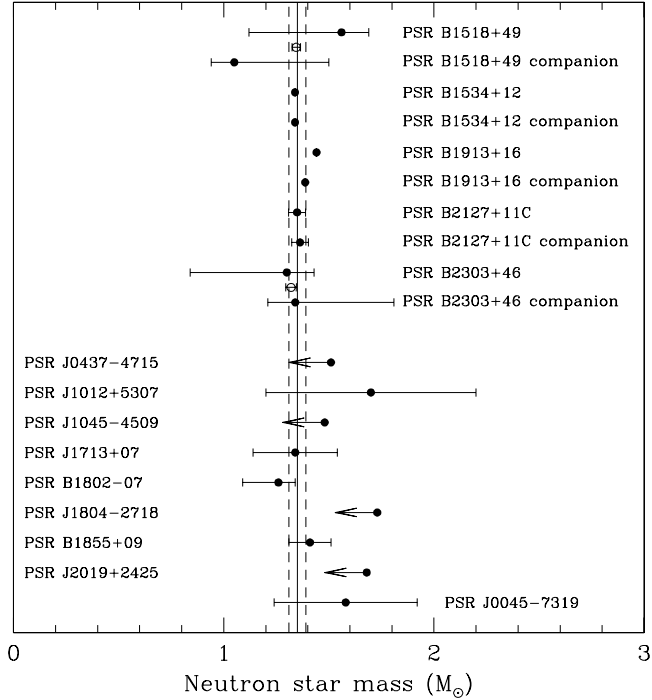}
\caption{
Neutron star masses from observations of close binaries
containing  radio pulsars.
 Error bars indicate 68\% ($1\sigma$) confidence limits. Upper limits,
indicated by arrows, are one-sided 95\% ($2\sigma$) confidence limits. Estimates of
masses of eight neutron stars in four double neutron star systems  are shown at the
top of the diagram. In two cases, the average neutron star masses
 in a system is known with
much better accuracy than the individual masses; these average masses are indicated
with open circles. Estimates of nine neutron stars in neutron star--white dwarf
binaries are shown in the central and bottom part of the diagram. The estimate  of
 of the neutron-star mass  in a neutron star--main-sequence star binary  is shown at the
 bottom of the diagram. Vertical lines are drawn at $M=1.35\pm 0.04~{\rm
M}_\odot$, and delimit  the 68\% bounds resulting from the  the maximum likehood
Gaussian distribution  of the measured neutron-star masses,  with mean mass
$1.35~{\rm M}_\odot$ and standard deviation $0.04~{\rm M}_\odot$.
From Thorsett and Chakrabarty (1999).
 }
\label{fig:NSmasses}
\end{center}
\end{figure}

The accuracy of measuring masses
of X-ray pulsars or
X-ray bursters is rather poor because
of many obstacles in obtaining high-precision
X-ray data, difficulties in establishing the parameters
of the orbital motions in the binary  and accounting for
many interfering factors (e.g., accretion flows, tidal forces).
Direct mass estimates have been obtained for several X-ray
pulsars and are displayed in Fig.\ \ref{fig:NSmasses}.
They will be hereafter denoted as $M_X$.

In what follows, we focus our attention at the binaries with
highest $M_X$.
Recent analysis  of the mass of neutron star in an X-ray binary,
Cyg X-2 
(which is an X-ray burster),
is -- at the $1\sigma$ (68\%) confidence level
$M_X(1\sigma)=1.78\pm 0.23~{\rm M}_\odot$. Simultaneously determined
 mass of the companion star
 in Cyg X-2 is
$M_{\rm c}=0.60\pm 0.13~{\rm M}_\odot$ (Orosz
and Kuulkers 1999). However, in order to use the  value of $M_X$
 to constrain dense matter EOS, it seems  reasonable to {\it require}
at least $2\sigma$ (i.e., 95\%) confidence level, which corresponds to
$M_X(2\sigma)=1.78\pm 0.46~M_\odot$. This would generate
$1.32~M_\odot$ as a lower limit on $M_{\rm max}$, which is not
 really useful
for constraining modern EOSs of dense matter. With additional constraint
on  the companion mass, resulting from theoretical models
($M_{\rm c}>0.75~{\rm M}_\odot$),  Casares et al. (1998) get
$M_X>1.88~{\rm M}_\odot$. This would be very restrictive as
far as the EOS is concerned,
but, in view of additional assumptions and
strong model dependence, it  cannot be used as
a clean  measurement of neutron star mass.

Another X-ray binary,
studied for  a long time, is Vela X-1 
(X-ray pulsar). 
The central value of $M_X$
obtained by different authors is  high, $\sim 2~{\rm M}_\odot$,
but the errors are unfortunately quite large.  In view of this,
the {\it lower bound}  on the mass of Vela X-1 is never significantly
higher than $1.4~{\rm M}_\odot$. Very recently, Barziv et al. (2001)
obtained  $M_X(2\sigma)>1.54~{\rm M}_\odot$ 
and $M_X(3\sigma)>1.43~{\rm M}_\odot$.
The $3\sigma$ lower bound has (formally) similar significance  as the binary --
radio-pulsar masses, but is unfortunately close  to the mass of the Hulse-Taylor
pulsar, and therefore does not yield a new constraint.

Other examples of
determination of neutron-star masses of the X-ray pulsars
  can be found in the paper of van Kerkwijk et al. (1995).
One has to conclude, that lower limits on $M_{\rm max}$,
determined at the $2\sigma$ confidence level from the condition
$M_{\rm max}>M_X$, are at present  too low to yield useful constraint on the
modern dense matter EOS.
This situation may change in the future. Very recently,
 Clark et al. (2002)
analysed   the high-mass X-ray binary 4U1700-37 
(in which the compact star is neither an X-ray pulsar 
nor an X-ray burster) 
and found
$M_X=2.44\pm 0.27~{\rm M}_\odot$, and $M_X(2\sigma)>2.0~{\rm M}_\odot$.
If the compact object {\it is a neutron star} then its mass would
rule out soft and moderately stiff EOSs (in particular EOSs
which include hyperons and phase transitions).

Observations of kHz quasi-periodic oscillations (QPOs)
 in low-mass X-ray binaries  could  (potentially)
yield  interesting constraints on neutron-star masses in
these binary systems, if this phenomenon is connected with the
orbital motion of matter around neutron star.
 However, as by the time of this
writing (July 2002), the  very mechanism  of the  phenomenon
 of the kHz QPOs is not established. Therefore,
 resulting constraints on neutron-star masses are
ambiguous.

Fortunately, there are close  binaries containing a radio pulsar
and a compact companion (another neutron star or a white dwarf).
The presence of a radio
pulsar has a great advantage over purely X-ray sources since
radio observations enable one to measure the pulsar
periods $P$ with extremely high precision (more than  12
  correct digits!).
Pulsars are known to be excellent timers: their proper
rotational periods (in comoving reference frames), in many cases,
are wonderfully stable. If, however, a pulsar participates in
orbital motion, its period, as measured  by a distant
observer, varies due to the Doppler effect. Actually,
 the radio
pulses  observed on Earth are also influenced  by the
spacetime curvature in the vicinity of the companion star.
Measurable relativistic effects include periastron advance, gravitational
redshift and transverse Doppler shift in the orbit, and  orbital decay
due to the emission of the gravitational radiation. Under favorable geometrical
conditions, Shapiro delay and geodetic precession can also be measured.
A review of the present status of the
timing analysis of the binary pulsars is  given in
a {\it Living Review} by Lorimer (2001).

The masses of six  neutron stars in the three
neutron-star binaries containing
 radio pulsars have been determined with high
accuracy. In the case of  PSR B1534+12,
 PSR B1913+16 and their neutron-star  companions, the masses
are known  within  $0.2\%$ and $0.02\%$, respectively.
Unfortunately, such  double neutron-star systems
are expected to be formed in a very specific evolutionary
scenario, and therefore  the most precisely measured 
neutron-star masses most probably do not give much information
about the complete neutron-star mass function.

An EOS of dense matter has to explain the  measured neutron-star masses.
A simple rule is to exclude those EOSs  which predict lower $M_{\rm max}$
 than the highest
{\it precisely measured}  neutron star mass
(in July 2002: 1.44 ${\rm M}_\odot$; in our opinion
result of
 Clark et al. (2002) has to be carefully
 checked as far as its  model-dependence is concerned
before being used to rule out soft and moderately stiff EOSs)
At present, this condition
excludes  only the softest  EOSs appearing in literature
(some EOSs  of dense matter with hyperons: see, e.g.,
Pandharipande 1971, Balberg and Gal 1997,
Vida{${\rm \widetilde n}$a et al. 2000).
 All of recently developed
EOS considered in the present
paper are consistent with the Hulse-Talor pulsar mass (see
Fig. \ref{fig:str-Mrhoc}).

Needless to say, it would be
highly desirable to accurately measure  the masses of more massive
neutron stars, the more massive the better. The highest
measured neutron star
mass $M^{\rm (max)}_{\rm obs}$  implies  the observational constraint
\begin{equation}
M_{\rm max}({\rm EOS})>M^{\rm (max)}_{\rm obs}~,
\label{boundMmax.obs}
\end{equation}
where $M_{\rm max}({\rm EOS})$ is maximum allowable mass for static
neutron star models for an EOS. 
Rotation at  periods  longer than say 5 ms
has almost no effect
 on $M_{\rm max}$.
A definite discovery of
$(1.8-2.0)~ M_\odot$ neutron star would rule out
EOSs
with hyperons and/or phase transitions;
the candidates for such massive neutron stars are  Vela X-1 and
 compact star in 4U1700-37,  discussed previously in this section.
A  discovery of, say,
2.1 $M_\odot$ neutron star would leave us with very stiff
EOSs  of dense matter containing nucleons only.
\section{Concluding  remarks}
\label{sect:final}
A theorist  who strongly believes in the power of the theory could find
a loophole in the criterion expressed by Eq.\ (\ref{boundMmax.obs}).
Why not imagine two families of ``neutron stars'', described by different
EOSs which correspond to two different forms (phases) of dense matter.
Let the first of them  has maximum mass $M_{\rm max}^{(1)}$ and the second one
$M_{\rm max}^{(2)}$. So one can can contemplate a situation, in which the first
EOS violates criterion (\ref{boundMmax.obs}) but is not eliminated by it because
the most massive observed ``neutron star'' belongs to a second family (Haensel
and Zdunik 1989).
For example, one might in  principle consider the existence  of strange quark
stars  {\it and} of neutron stars, or even of Q-stars {\it and} neutron stars.
Of course, a necessary condition for some contact with reality is
the stability of {\it both}  these families. This means, that at the same baryon
numbers compact star configurations
 belonging to different families have to be separated
by a sufficiently high energy barrier.

How far can one go with theoretical considerations not
substantiated by solid experimental basis? Many theorists (to
which the author belongs) think that respecting  the {\it Occam's
razor}  principle is necessary  in research
work.\footnote[8]{Occam's razor:  a principle stating that
entities must not be multiplied beyond what is necessary; often
interpreted to mean that phenomena should be explained in terms of
the simplest possible causes. William of Occam was a medieval
English philosopher who devised this principle.}
The fact that a model is not ruled out by observations is not a proof of its
reality. The model has to be {\it necessary}  for understanding the
observations to be considered as representing
a reality.
It would be   intellectually  arrogant  to believe that  the
universe is filled with  objects predicted by the theories based on
a distant  extrapolation of the laboratory physics.
Conventional  models of neutron stars should be considered  as long as they are
sufficient to understand astrophysical phenomena and measurements.
If one day a compact object
of, say, $8~{\rm M}_\odot$ is discovered which {\it is not} a black
hole,   then it cannot be but a Q-star. Future observations will hopefully bring
us more information, and enrich our knowledge of the compact stars,
but for the time being the principle devised by  William
of Occam  is a useful complement to the theory of dense matter.
\parindent 0pt
\vskip 3mm
{\it Acknowledgements.} The author is deeply grateful to D.G. Yakovlev for reading
the manuscript and for helpful remarks and comments.
He is also very grateful to A.Y. Potekhin
for his precious help in the preparation of figures. This work was supported by
the KBN grant No. 5 P03D 020 20.

\end{document}